\documentclass[useAMS,usenatbib]{mn2e}
\usepackage{graphicx}  
\usepackage{ amssymb }

\def\simgt{\lower.5ex\hbox{$\; \buildrel > \over \sim \;$}}
\def\simlt{\lower.5ex\hbox{$\; \buildrel < \over \sim \;$}}

\def\msun{M$_\odot$}
\def\Teff{$T_{\rm eff}$}
\def\teff{$T_{\rm eff}$}

\def\alfa2d{MLT--$\,\alpha^{\rm 2D}$}

\def\Dnu{$\Delta\nu$}
\def\numax{$\nu_{\rm max}$}

\title[]{Asteroseismology of old open clusters with {\it Kepler}\,: \\direct estimate of the integrated RGB mass loss in NGC6791 and NGC6819}

\author[A. Miglio et al.]{A. Miglio$^{1}$\thanks{E-mail: miglioa@bison.ph.bham.ac.uk}, K. Brogaard$^{2}$, D. Stello$^{3}$, W. J. Chaplin$^{1}$, F. D'Antona$^{4}$, J. Montalb{\'a}n$^{5}$, 
\newauthor S. Basu$^{6}$, A. Bressan$^{7,8}$,  F. Grundahl$^{9}$,  M. Pinsonneault$^{10}$, A. M. Serenelli$^{11}$, 
\newauthor Y. Elsworth$^{1}$, S. Hekker$^{12,1}$,  T. Kallinger$^{13}$, B. Mosser$^{14}$, P. Ventura$^{4}$,  A. Bonanno$^{15}$,
\newauthor A. Noels$^{5}$,  V. Silva-Aguirre$^{16}$, R. Szabo$^{17}$, J. Li$^{18}$, S. McCauliff$^{19}$, C. K. Middour$^{19}$,
\newauthor  H. Kjeldsen$^{9}$\\
$^{1}$School of Physics and Astronomy, University of Birmingham, Edgbaston, Birmingham B15 2TT, United Kingdom\\
$^{2}$Department of Physics and Astronomy, University of Victoria, P.O. Box 3055, Victoria, BC V8W 3P6, Canada\\
$^{3}$Sydney Institute for Astronomy (SIfA), School of Physics, University of Sydney, NSW 2006, Australia\\
$^{4}$INAF-Osservatorio Astronomico di Roma, via Frascati 33, Monteporzio Catone (RM), Italy\\
$^{5}$Institut d'Astrophysique et de G\'eophysique de l'Universit\'e de Li\`ege, All\'ee du 6 Ao\^ut, 17 B-4000 Li\`ege, Belgium\\
$^{6}$Department of Astronomy, Yale University, P.O. Box 208101, New Haven CT 06520-8101, USA\\
$^{7}$SISSA, via Bonomea, 265, 34136 Trieste, Italy\\
$^{8}$INAF-Osservatorio Astronomico di Padova, Vicolo dell'Osservatorio 5, I-35122 Padova, Italy\\
$^{9}$Department of Physics and Astronomy, Building 1520, Aarhus University, 8000 Aarhus C, Denmark\\
$^{10}$Ohio State University, Dept. of Astronomy 140 W. 18th Ave. Columbus, OH 43210 USA\\
$^{11}$Institute for Space Sciences (CSIC-IEEC), Facultad de Ciencies, Campus UAB, 08193 Bellaterra, Spain\\
$^{12}$Astronomical Institute ÔAnton PannekoekÕ, University of Amsterdam, Science Park 904, 1098 XH Amsterdam, The Netherlands\\
$^{13}$Department of Physics and Astronomy, University of British Colombia, 6224 Agricultural Road, Vancouver, BC V6T 1Z1, Canada\\
$^{14}$LESIA, CNRS, Universit{\'e} Pierre et Marie Curie, Universit{\'e} Denis Diderot, Observatoire de Paris, 92195 Meudon cedex, France\\
$^{15}$INAF-Osservatorio Astrofisico di Catania, Via S.Sofia 78, 95123, Catania, Italy\\
$^{16}$Max Planck Institute for Astrophysics, Karl-Schwarzschild-Str. 1, 85748, Garching bei M\"unchen, Germany\\
$^{17}$Konkoly Observatory of the Hungarian Academy of Sciences, Konkoly Thege Mikl\'os \'ut 15-17, H-1121 Budapest, Hungary\\
$^{18}$SETI Institute/NASA Ames Research Center, Moffett Field, CA 94035\\
$^{19}$Orbital Sciences Corporation/NASA Ames Research Center, Moffett Field, CA 94035\\
}         
\begin{document}
\date{}
\pagerange{\pageref{firstpage}--\pageref{lastpage}} \pubyear{2010}

\maketitle


\begin{abstract}
Mass loss of red giant branch (RGB) stars is still poorly determined
, despite its crucial role in the chemical enrichment of galaxies. Thanks to the recent detection of solar--like oscillations in G--K giants in open clusters with $Kepler$, we can now directly determine stellar masses for a statistically significant sample of stars in the old open clusters NGC6791 and NGC6819. The aim of this work is to 
constrain the integrated RGB mass loss by comparing the average mass of stars in the red clump (RC) with that of stars in the  low-luminosity portion of the RGB (i.e. stars with $L\lesssim L({\rm RC})$).  Stellar masses were determined by combining the available seismic parameters \numax\ and \Dnu\  with additional photometric constraints and with independent distance estimates. We measured the masses of 40 stars on the RGB and 19 in the RC of the old metal-rich cluster NGC6791. We find that the difference between the average mass of RGB and RC stars  is small, but significant ($\Delta \overline{M}=0.09\pm 0.03$ (random) $\pm 0.04$ (systematic) M$_\odot$). Interestingly, such a small $\Delta\overline{M}$ does not support scenarios of an extreme mass loss for this metal-rich cluster. If we describe the mass-loss rate with Reimers' prescription, a first comparison with isochrones suggests that the observed $\Delta\overline M$ is compatible with a mass-loss efficiency parameter in the range $0.1 \lesssim \eta \lesssim 0.3$.   
Less stringent constraints on the RGB mass-loss rate are set by the analysis of the $\sim 2$ Gyr-old NGC6819, largely due to the lower mass loss expected for this  cluster, and to the lack of an independent and accurate distance determination. 
{ In the near future, additional constraints from frequencies of individual pulsation modes and spectroscopic effective temperatures, will allow further  stringent tests of the \Dnu\ and \numax\, scaling relations, which provide a novel, and potentially very accurate,  means of determining stellar radii and masses.}

\end{abstract}


\begin{figure*}
\begin{center}
\resizebox{.65\hsize}{!}{\includegraphics[]{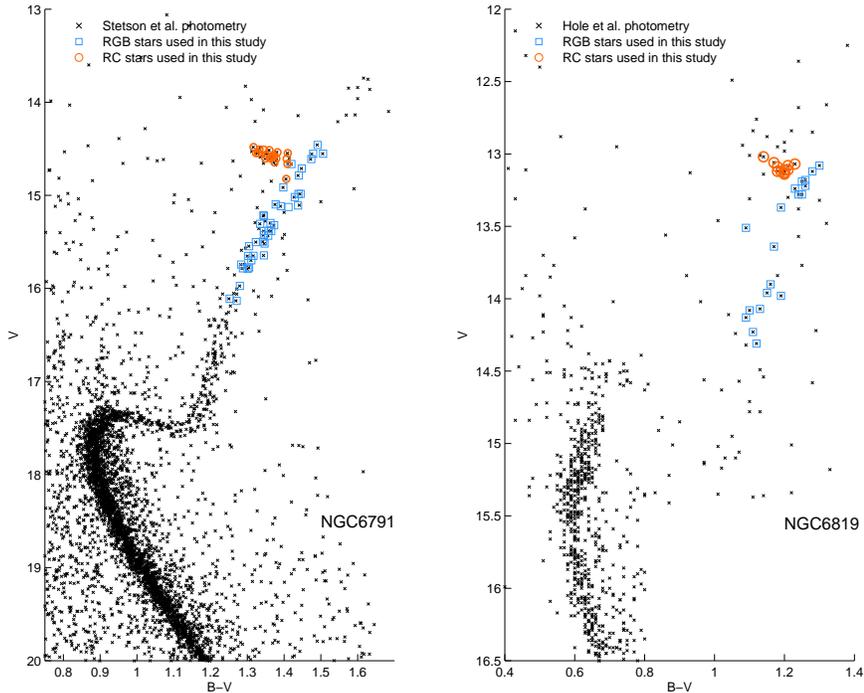}}
\caption{\small Colour--magnitude diagram of NGC6791 (left panel) and NGC6819 (right panel). Photometric data are taken from \citet{Stetson2003} and \citet{Hole2009}, respectively. RGB stars used in this work are marked by open squares and RC stars by open circles. See Sec. \ref{sec:data} for a description of the target selection. }
\label{fig:fotom}
\end{center}
\end{figure*}
\section{Introduction}
The mass-loss rates of evolved stars are of primary importance for
stellar and galactic evolution models, but neither the observations nor the
theory are adequate to place reliable direct quantitative constraints.
The mass loss from Asymptotic Giant Branch (AGB) stars is reasonably well understood in terms of
global pulsations lifting gas out to distances above the photosphere
where dust forms \citep[e.g.][]{Wood1979, Bowen1988}, and the action of
radiation pressure on the dust that further drives dust and gas away
\citep[e.g.][and references therein]{Hoefner2009}. In contrast,
there is no reliable theory for the mass loss of cool, dust-free red giants, and
even the wind acceleration mechanism remains unknown for the Red Giant Branch (RGB). 
For these stars no 
known mechanisms can yet
satisfactorily explain
the observed wind characteristics \citep[for further discussion,
see][]{Lafon1991, Harper1996}.

Indirect information is, however, available on the integrated mass loss on the first
giant branch \citep[see e.g.][for a recent review]{Catelan09}. This is important because it may involve a substantial amount of mass, especially for the lower mass stars undergoing
the helium core flash, and thus may affect the initial to final mass
relation and the amount of mass recycled into the interstellar medium.
In the case of Globular Cluster (GC) stars, mass loss along the RGB,
and its spread, affect the distribution
of stars along the Horizontal Branch (HB). This is because the morphology of the core helium burning track
depends on the ratio between the helium core mass and the hydrogen--rich envelope
mass, or, in other words, on the relative roles played by the core helium burning and the
hydrogen shell burning \citep[e.g.][]{Sweigart1976}. The indirect evidence
tells us that in GCs more mass is lost on the RGB than the AGB,
since the typical stellar mass on the HB is $\sim$0.6 \msun,
 with turn-off masses of $\sim$0.8 \msun, and white dwarfs masses of
$\sim$0.5--0.55~\msun\ \citep{Kalirai2009}.

The standard picture that the mass
distribution along the HB matches mass loss on the RGB has been questioned in the last decade. 
The ubiquitous presence of multiple populations
in GCs has now become evident \citep[e.g.][]{Gratton2001, Carretta2009, Piotto2007},
and the proposal that
the stellar helium content of the different populations also has a role in determining
the HB morphology \citep{D'Antona2002, D'Antona2004} was confirmed by the presence of multiple
main sequences in some globular clusters \citep{Piotto2005, Piotto2007}

In order to describe the mass loss along the RGB, researchers rely on empirical
 laws like that of \citet{Reimers1975a, Reimers1975b}, which is based on observations of Population I giants, and which describes mass-loss rates as a function of stellar luminosity, radius, and surface gravity. While subsequent work \citep{Mullan1978, Goldberg1979, Judge1991,
Catelan2000, Schroder2007} led to slight refinements, a variant of the
``Reimers' law" is generally used to compute stellar evolution 
models of cool stars at all ages and metallicities.

Several observational indications on the RGB mass loss have emerged in recent years.
\cite{Origlia2002} detected the circumstellar matter around GC
red giants using the ISOCAM camera onboard the ISO satellite, and derived
mass-loss rates for a few RGB stars. A first empirical mass-loss
law was proposed by \cite{origlia2007}, based on the observations by IRAC, onboard Spitzer,
of the GC 47~Tuc. They found mass loss to be episodic, 
depending less on luminosity than in Reimers' law, and occurring predominantly in the upper
$\sim$2~mag of the RGB \citep{origlia2010}. These findings were questioned by \citet{Boyer2010} and \citet{McDonald2011a,McDonald2011b}, who found no evidence for dust production in stars with $L < 1000$ L$_\odot$. By comparing with HB models they showed that mass loss on the RGB of 47~Tuc is likely to be smaller than 0.24 M$_\odot$.
{ Constraints on the dust production and mass loss are now becoming available for giants belonging to other GCs   (\citealt{McDonald2009}, \citealt{Boyer2009}, \citealt{McDonald2011c}, and the survey by Origlia et al.), and will eventually provide a complete investigation of the differential mass loss among clusters with different metallicities and HB morphologies. }

Asteroseismology promises to shed new light on this problem by directly measuring the masses of giant stars in open clusters. The NASA $Kepler$ mission, which was launched successfully in March 2009, includes in its field of view two old open clusters, NGC6791 and NGC6819. In these clusters solar-like oscillations were detected in about 100 giants (see \citealt{Stello2010, Basu2011, Hekker2011} and Fig. \ref{fig:fotom}), providing independent constraints on the masses and radii of stars on the RGB, as well as on the distance to the clusters and their ages \citep{Basu2011}.

The goal of this paper is to constrain the RGB mass loss by comparing the average mass of stars in the red clump (RC) with  the  low-luminosity portion of the RGB (i.e. with $L\lesssim L({\rm RC})$).  The structure of the paper is as follows.
In Sec. \ref{sec:mass} we describe the procedures used to estimate the masses of stars from the available seismic and non-seismic constraints. We then address the specific case of NGC6791 in Sec. \ref{sec:6791}, and of NGC6819 in Sec. \ref{sec:6819}. A first comparison with model predictions is presented in Sec. \ref{sec:massloss}, while a brief summary and future prospects are given in Sec. \ref{sec:summary}.

\section{Estimating stellar masses}
\label{sec:mass}
To estimate stellar masses we use the average seismic parameters that characterise solar-like oscillation spectra: the so-called average large frequency separation (\Dnu), and the frequency corresponding to the maximum observed oscillation power (\numax).
The large frequency spacing is expected to scale as the square root of the mean density of the star:
\begin{equation}
\Delta\nu\simeq\sqrt{\frac{M/{\rm M}_\odot}{(R/{\rm R}_\odot)^3}}\Delta\nu_{\odot}{\rm,}
\label{eq:dnu}
\end{equation}
where $\Delta\nu_\odot=135$ $\mu$Hz. The frequency of maximum power is approximatively proportional to the acoustic cutoff frequency \citep{Brown1991, Kjeldsen1995, Mosser2010, Belkacem2011}, and therefore:
\begin{equation}
\nu_{\rm max}\simeq\frac{M/{\rm M}_\odot}{(R/{\rm R}_\odot)^2\sqrt{T_{\rm eff}/{\rm T}_{\rm eff,\odot}}}\nu_{\rm max,\odot}\,{\rm,}
\label{eq:numax}
\end{equation}
where $\nu_{\rm max,\odot}=3100$ $\mu$Hz and ${\rm T}_{\rm eff,\odot}=5777$ K.

These scaling relations are widely used to estimate masses and radii of red giants \citep[see e.g.][]{Stello2008,Kallinger2010, Mosser2010}, but they are based on simplifying assumptions that remain to be independently verified. Recent advances have been made on providing a theoretical basis for the relation between the acoustic cut-off frequency and \numax\  \citep{Belkacem2011}, and preliminary investigations with stellar models \citep{Stello2009a} indicate that the scaling relations hold to within $\sim3$\% on the main sequence and RGB \citep[see also][]{White2011}. 

Depending on the observational constraints available,  we may derive mass estimates from Equations \ref{eq:dnu} and \ref{eq:numax} alone, or via their combination with other available information from non-seismic observations. 
When no information on distance/luminosity is available, which is usually the case for field stars, Eq. \ref{eq:dnu} and \ref{eq:numax} may be solved to derive $M$ and $R$ \citep[see e.g.][]{Kallinger2010, Mosser2010}:
\begin{eqnarray}
\frac{M}{{\rm M}_\odot} &\simeq& \left(\frac{\nu_{\rm max}}{\nu_{\rm max, \odot}}\right)^{3}\left(\frac{\Delta\nu}{\Delta\nu_{\odot}}\right)^{-4}\left(\frac{T_{\rm eff}}{{\rm T}_{\rm eff, \odot}}\right)^{3/2} \label{eq:scalM}      \\
\frac{R}{{\rm R}_\odot} &\simeq& \left(\frac{\nu_{\rm max}}{\nu_{\rm max, \odot}}\right)\left(\frac{\Delta\nu}{\Delta\nu_{\odot}}\right)^{-2}\left(\frac{T_{\rm eff}}{{\rm T}_{\rm eff, \odot}}\right)^{1/2}{\rm.}\label{eq:scalR}
\end{eqnarray}

In the case of clusters, we can use the distance/luminosities estimated with independent methods (i.e. via isochrone fitting or eclipsing binaries) as an additional constraint. Including this information allows $M$ to be estimated also from Eq. \ref{eq:dnu} or \ref{eq:numax} alone (see Eq. \ref{eq:scalM2} and \ref{eq:scalM3}, respectively), or in combination leading to a mass estimate with no explicit dependence on \Teff\ (as in Eq. \ref{eq:scalM4}):

\begin{eqnarray}
\frac{M}{{\rm M}_\odot} &\simeq& \left(\frac{\Delta\nu}{\Delta\nu_{\odot}}\right)^{2}\left(\frac{L}{{\rm L}_\odot}\right)^{3/2}   \left(\frac{T_{\rm eff}}{{\rm T}_{\rm eff, \odot}}\right)^{-6} \label{eq:scalM2}      \\
\frac{M}{{\rm M}_\odot} &\simeq& \left(\frac{\nu_{\rm max}}{\nu_{\rm max, \odot}}\right)\left(\frac{L}{{\rm L}_{\odot}}\right)\left(\frac{T_{\rm eff}}{{\rm T}_{\rm eff, \odot}}\right)^{-7/2}  \label{eq:scalM3}     \\
\frac{M}{{\rm M}_\odot} &\simeq& \left(\frac{\nu_{\rm max}}{\nu_{\rm max, \odot}}\right)^{12/5}\left(\frac{\Delta\nu}{\Delta\nu_{\odot}}\right)^{-14/5}\left(\frac{L}{{\rm L}_\odot}\right)^{3/10} {\rm .}\label{eq:scalM4}    
\end{eqnarray}

In the following sections we use Equations \ref{eq:scalM} to \ref{eq:scalM4} directly to estimate $M$ (and $R$) without adding any extra dependence on stellar models. As illustrated in detail e.g. by \citet{Gai2011}, additional (so-called ``grid-based'') methods to estimate $M$ and $R$ can be designed. These procedures are also based on  Eqs. \ref{eq:dnu} and \ref{eq:numax} but, by searching solutions for $M$ and $R$ in grids of evolutionary tracks, have the advantage of reducing uncertainties on the derived mass and radius, at the price of some model dependence that we prefer to avoid in this study. 
 \subsection{Error estimates}
 \label{sec:errors}
The formal uncertainties ($\sigma_i$) on the masses ($M_i$) of the stars were used to compute a weighted average mass for stars belonging to the RGB and for stars in the RC:
$$
\overline{M}=\frac{\sum_1^N{M_i/\sigma_i^2}}{\sum_1^N{1/\sigma_i^2}}.
$$
The uncertainties in these averages were estimated from the weighted scatter in the masses, i.e.
$$
\sigma_{\overline{M}}=t[N-1]\,\sqrt{{\frac{\sum_1^N{\frac{\left(M_i-\overline{M}\right)^2}{\sigma_i^2}}}{(N-1)\sum_1^N1/\sigma_i^2}}}.
$$
Due to the small number of stars in the samples, the correction factor $t[N-1]$ was drawn from  the Student's $t$ distribution with $N-1$ degrees of freedom in order to assess the confidence limits correctly \citep[see e.g.][]{Chaplin1998}. We adopted 68\% as the 1-$\sigma$ confidence level.

To assess how well the formal fitting uncertainties reflected the scatter in the data we computed the ratio $z=\sigma_{\overline{M}}/\overline{\sigma_{\rm M}},$  where $\overline{\sigma_{\rm M}}$ is the weighted mean uncertainty estimated from the formal uncertainties on the masses, i.e. : $$\overline{\sigma_{\rm M}}=\left(\sum_1^N{1/\sigma_i^2}\right)^{-1/2}{\rm.}$$
If mass scatter is dominated by random errors and the formal uncertainties reflect the true observational uncertainties, then $z\sim1$. 

The uncertainty on $\Delta\overline{M}=\overline{M}_{\rm RGB}-\overline{M}_{\rm RC} $ due to random errors was finally computed as:
$$
\sigma_{\Delta \overline{M}}=\sqrt{\sigma_{ \overline{M}_{\rm RGB}}^2+\sigma_{ \overline{M}_{\rm RC}}^2}.
$$

\section{NGC 6791}
\label{sec:6791}
NGC6791 is an old open cluster, with age estimates ranging from  $\sim$7 to $\sim$12 Gyr, depending on the
choice of reddening and metallicity \citep[see][]{Chaboyer1999, Stetson2003}. More recent studies \citep{Brogaard2011, Basu2011} point towards the lower end of this age interval. This cluster has a solar [$\alpha$/Fe] ratio \citep{Origlia2006}
and is one of the most metal-rich clusters in our Galaxy \citep{Friel1993} with a metallicity of 
[Fe/H]=+0.3 to +0.5 \citep[e.g.][and references therein]{Brogaard2011,Origlia2006}.

\subsection{Mass loss in NGC6791: a puzzling open question}
\label{sec:mloss6791}
As summarised in the following paragraphs, an issue widely debated in the literature is whether the high metallicity of NGC6791 may lead to higher-than-average mass-loss
rates along the RGB, and whether such enhanced mass loss could explain the very peculiar luminosity distribution of its white dwarfs. 

By examining the white dwarf data in the deep HST photometry analysed by \cite{king2005},
\cite{bedin2005} found a rich population of hundreds of white dwarfs.
Surprisingly, the luminosity function of these stars showed two peaks, the
dimmer one possibly consistent with the ``normal" turnoff age of $\sim$ 8 Gyr
(but see below), the brighter one
indicating a much younger age of $\sim$ 2.4 Gyr. White dwarf dating of open and
globular clusters has often been used in the literature, and generally provides
results consistent with the turnoff dating \citep{vonhippel2005,hansen2007}.
For NGC 6791, \cite{hansen2005} speculated that
 the peculiar luminosity function could be related to high rates of mass loss. If all the cluster stars
formed at the same epoch ($\sim$ 8 Gyr ago), and if there has been significant
mass loss on the red giant branch, many stars may have been reduced to a
mass below the threshold to ignite the core
helium flash. They would evolve much more slowly than the normal carbon--oxygen
white dwarfs in the cluster, leading to a more luminous peak in the white dwarf
luminosity function.
This explanation proposed by \cite{hansen2005} received support from \cite{kalirai2007}, who derived a very low average mass (M$\sim$0.43$\pm$0.06 \msun) for the luminous white dwarfs for which they obtained Keck/LRIS spectra,
 well below the limit for helium ignition in a degenerate core.
In addition, the post-helium flash, helium-core-burning stars
in this cluster are divided into a red giant clump and an extremely blue HB
\citep[e.g.][]{kalirai2007}. The latter may be populated by the less massive
stars evolving from the RGB that were still able to ignite a late helium core flash
\citep[e.g.][]{castellani1993,brown2001}. Stars with slightly
higher mass loss do not ignite helium and evolved directly
into helium white dwarfs, giving rise to
the more luminous peak in the white dwarf luminosity function.
Since the mass at the turnoff of NGC 6791 is $\sim$1.1 \msun
\citep{Brogaard2011}, a huge mass loss, possibly connected to the high metallicity of the cluster, must have occurred if the above scenario is true.

This tentative explanation was disputed by \citet{vanLoon2008}, who found that the
RGB luminosity function does not show any signs of star depletion, and that the number of
HB (clump) stars is consistent with most cluster stars evolving through the
helium flash. In addition,  dust -- a sign of extreme mass loss --
was not detected around the most luminous giants.
More recently, \cite{Bedin2008} used additional data to examine the white dwarf
sequence in NGC 6791, confirming a second dimmer peak
in the luminosity function.
Since the first peak is $\sim$0.75~mag brighter than the dimmer one,
\cite{bedin2008a} proposed that binary white dwarfs are responsible for the
brighter peak. This explanation required that more than $\sim$30\% of the white dwarfs are binaries. 
In this context, the extreme HB can be attributed to binary evolution following
the same scenario as the production of extreme B and O subdwarfs in the Galactic
field \citep[e.g.][]{han2003}.

In the new data analysed by \cite{bedin2008a}, a
helium white dwarf population does not seem to agree with the
colour distribution of the cooling sequences, while
a further problem emerged for the interpretation of the dimmer peak, in that its
cooling age is $\sim$6 Gyr, compared with the $\sim$8 Gyr turnoff age.
This latter problem has been recently addressed by including in the modelling the energy released by the sedimentation
of $^{22}$Ne and of the C--O mixture in the white dwarf core
\citep[][]{deloye2002,garcia-berro2010,althaus2010}, which has the effect of lengthening the evolution of the C--O white dwarfs.
The recent analysis by \citet{garcia-berro2011} also favours a significant population of unresolved binaries as the most likely explanation for the observed white-dwarf luminosity function.

\begin{table*}
 \begin{tabular}{lcccccc}
\hline
Eq. &  $\overline{M}_{\rm RGB}\pm\sigma_{\overline{M}}$ & $z_{\rm RGB}$ & $\overline{M}_{\rm RC}\pm\sigma_{\overline{M}}$ & $z_{\rm RC}$ & $\Delta \overline{M}\pm\sigma_{\Delta\overline{M}}$  \\
\hline
 (\ref{eq:scalM}) &     $1.23 \pm     0.02$ &      1.8  &    $1.03 \pm     0.03 $   &  1.9  &    $0.19  \pm    0.04$\\ 
 (\ref{eq:scalM2}) &   $  1.22 \pm      0.01\ (0.10)$&      0.7&     $ 1.20\pm       0.02\ (0.10)$  &    0.6 &  $   0.02\pm       0.02\ (0.01)$\\ 
 (\ref{eq:scalM3}) &   $  1.23  \pm     0.01\ (0.07)$&      1.1  &   $ 1.14\pm       0.02\ (0.06)$  &    1.1  &  $  0.09\pm       0.02\ (<0.01)$ \\
 (\ref{eq:scalM4}) &   $  1.23  \pm     0.02\ (0.02)$&      1.9  &    $1.07\pm       0.03\ (0.02)$ &    2.0  &   $ 0.16 \pm      0.03\ (<0.01)$ \\
\hline
 \end{tabular}
\caption{\small NGC6791: average mass of stars in the RGB and RC estimated using different observational constraints and scaling relations (Equations \ref{eq:scalM}, \ref{eq:scalM2},
\ref{eq:scalM3}, and \ref{eq:scalM4}). The systematic error on $\overline M$ due to the uncertainty on the distance is reported in brackets. $z$  (see Sec. \ref{sec:errors}) was computed taking into account only random errors in the observables,  and the correlation between BC and \teff. No correction to the  \Dnu\ scaling was applied to clump stars.} \label{tab:6791Mnocorr}

 \begin{tabular}{lccc}
\hline
Eq. &  $\overline{M}_{\rm RC}\pm\sigma_{\overline{M}}$ & $z_{\rm RC}$ & $\Delta \overline{M}\pm\sigma_{\Delta\overline{M}}$  \\
\hline
 (\ref{eq:scalM}) &    $1.15 \pm     0.03 $   &  1.8  &    $0.08  \pm    0.04 $\\ 
 (\ref{eq:scalM2}) &  $ 1.14\pm       0.02\ (0.09) $  &    0.6 &  $   0.08\pm       0.02\ (0.01) $\\ 
 (\ref{eq:scalM3}) &  $ 1.14\pm       0.02\ (0.06)$  &    1.1  &  $  0.09\pm       0.02\ (<0.01) $ \\
 (\ref{eq:scalM4}) &  $1.15\pm       0.03\ (0.02)$ &    2.0  &   $ 0.08 \pm      0.03\ (<0.01) $ \\
\hline

 \end{tabular}
\caption{\small As in Table \ref{tab:6791Mnocorr},  after applying the 2.7\% suggested correction to the \Dnu\ scaling for clump stars (see Sec. \ref{sec:correction}).} \label{tab:6791Mcorr}
\end{table*}

\subsection{Targets and data}
\label{sec:data}

For our initial set of targets we took the stars listed by \citet{Stello2011b}, but removed those listed as seismic non-members and blended stars. 
To avoid possible contamination by AGB stars, we then removed stars with
$L>L({\rm RC})$, and subsequently separated the remaining stars into two
groups, RGB and RC stars, by visual inspection of the CMD.  This
classification agrees with the one presented by Stello et al. (2011a, Table
1).
Recent work by \citet{Montalban2010}, \citet{Bedding2011}, and \citet{Mosser2011} has shown that the characteristics of the $\ell=1$ ridge in
the {\'e}chelle diagram  can be used as an indicator of evolutionary state of low-mass giants. A detailed and systematic analysis of the
properties of $\ell=1$ modes in cluster stars is beyond the scope of this paper. Nonetheless, by visual inspection of
{\'e}chelle diagrams we could identify two misclassified stars, by looking at the typical spacing between $\ell=1$ modes
\citep[see examples in][]{Bedding2011}. The first one is KIC2437103, which we now identify as a core-He burning star, and
the second one is KIC2437589 which has a spectrum not compatible with a RC star, and deserves further analysis. 
After this selection, we are left with 40 stars on the RGB and 19 stars in the RC (see Fig. \ref{fig:fotom}).

If the mass loss is very high then the helium-burning stars that lost the most mass will not be in the RC, but rather on the corresponding flat part of the horizontal branch {at higher $T_{\rm eff}$ and lower $L$}. 
Since our sample only includes RC stars, one might suspect that this introduces a selection bias. However, as shown in the membership study by \citet{Platais2011}, there are only 12 potential low-mass helium-burning stars outside the RC (disregarding the EHB stars, which most likely result from binary evolution). Among these, many are likely to be blue stragglers or non-members (see their Fig. 1 and Table 1). Furthermore, work in progress (Brogaard et al. 2011, in prep.) shows that a theoretical ZAHB that matches the RC stars is brighter than the potential low-mass HB stars. It also shows that the bluest of these (star 2-17, first analysed by \citet{Peterson1998} and often associated with the HB), is actually a blue straggler. Based on this, we believe that there are no, or at most a couple, helium-burning stars present with masses smaller than those in the RC.

The photometric time-series data were obtained by the \textit{Kepler} space
telescope \citep{Koch2010} 
between 2009 May 12 and 2010 March 20, known as
observing quarters 1--4.  We used the  `long cadence' observing
mode ($\Delta t =29.4\,$min, \citealt{Jenkins2010a}), 
which provided about 14,000 data points per star.  The data reduction from
raw images to the final light curves is described by
\citet{Jenkins2010b}, \citet{Garcia2011}, and \citet{Stello2011b}. 
A detailed description of the asteroseismic parameters 
used in this work is given by \citet{Stello2011a}, who used the
method of \citet{Huber2009} to measure \Dnu\ and \numax.
The average uncertainties in \Dnu\ and \numax\ in this sample of targets are 1.3\% and 1.7\%,
respectively.  Other analyses of the same cluster are given in our companion
papers: \citet{Basu2011}, \citet{Hekker2011}, and \citet{Stello2011a}.

The effective temperatures used in this study are, following \citet{Basu2011}, \citet{Hekker2011}, \citet{Stello2011b}, and \citet{Stello2011a}, based on the
$(V-K)$ colour and on the calibrations of \citet{Ramirez2005}. The value we adopted for the reddening is
E$(B-V)=0.16\pm0.02$~mag \citep{Brogaard2011} and for the metallicity [Fe/H]=+0.3. As presented by \citet{Hekker2011}, the estimated random star-to-star error on
\Teff\ is about 50 K, which we also adopt in the following analysis. However,  systematic errors on the \Teff\ scale due to colour--temperature
calibrations and reddening could increase the error budget up to $\sim 110$ K \citep{Hekker2011}; we discuss
in Sec. \ref{sec:correction} the effect of systematic errors in \Teff\ on our results.

Finally, luminosities were derived using $V$ magnitudes from \citet{Stetson2003}, the bolometric corrections by
\citet{Flower1996}, and the distance modulus  $(m- M)_V=13.51 \pm  0.06$~mag estimated by \citet{Brogaard2011} using eclipsing binaries.

\begin{figure*}
\begin{center}
\resizebox{.85\hsize}{!}{\includegraphics[]{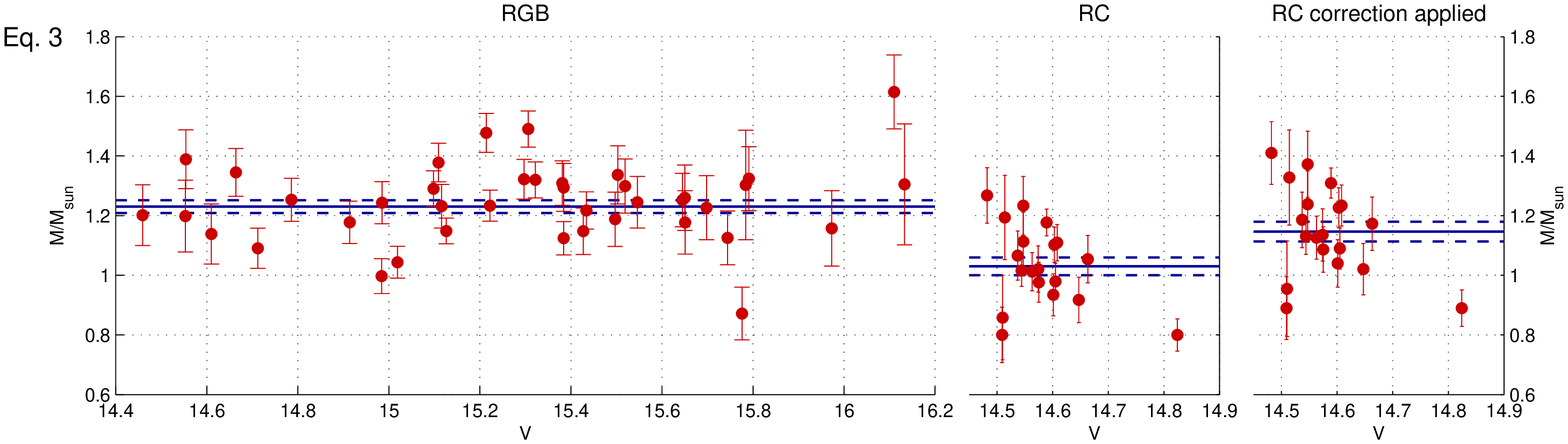}}
\resizebox{.85\hsize}{!}{\includegraphics[]{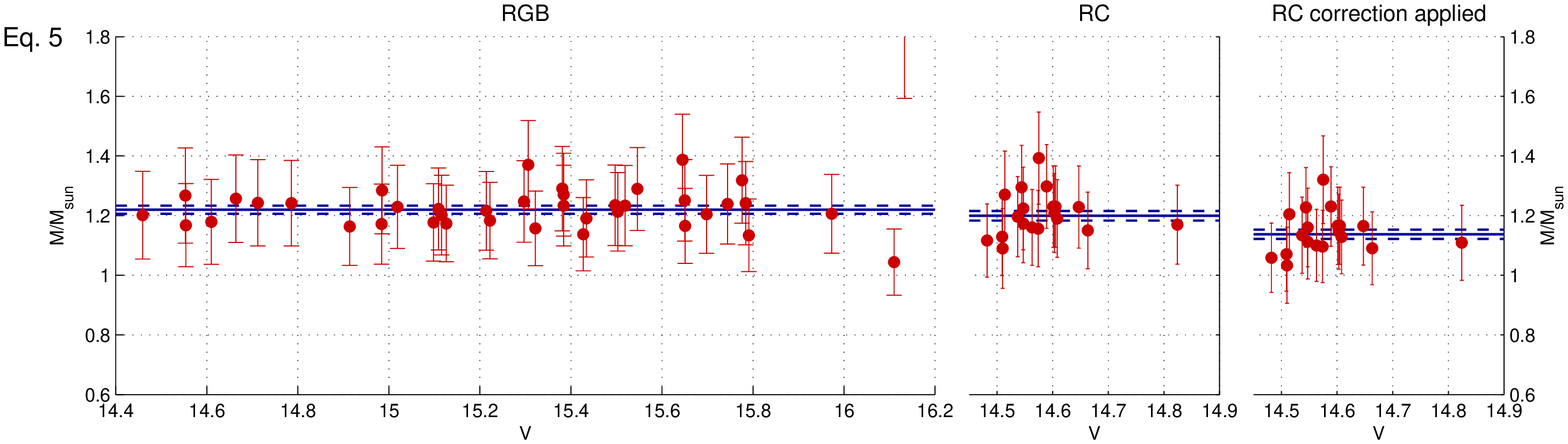}}
\resizebox{.85\hsize}{!}{\includegraphics[]{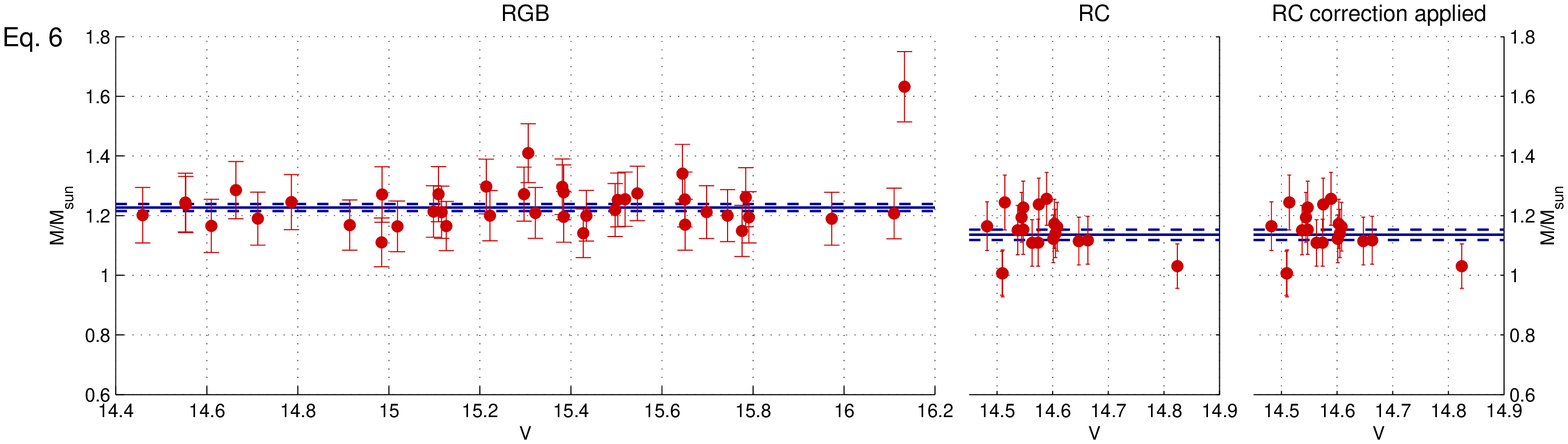}}
\resizebox{.85\hsize}{!}{\includegraphics[]{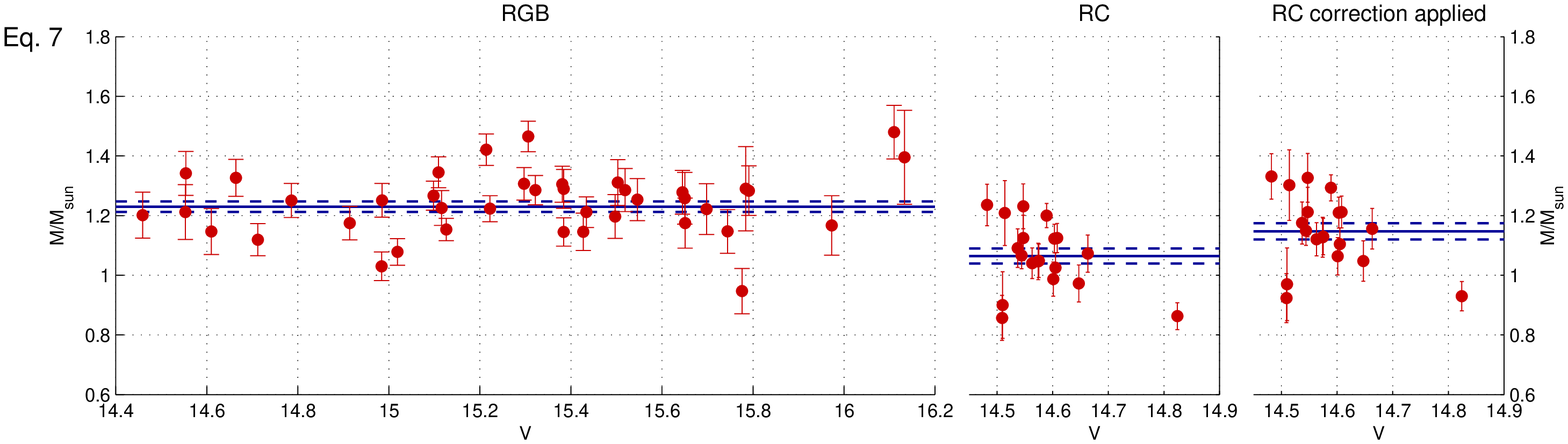}}

\vspace{0.5cm}
\caption{\small Mass of red giants in NGC6791 as a function of visual magnitude in the RGB (left panels) and in the RC (central panels). {Random errors on the mass estimate of each star  are represented by vertical error bars.} The average mass of RGB and RC stars is represented with full lines, and the 1-$\sigma$ error bars with dashed lines.  Right panels show the masses of RC stars after a correction to the \Dnu\ scaling is applied (see Sec. \ref{sec:correction}).}
\label{fig:masses6791}
\end{center}
\end{figure*}

\subsection{Results}
\label{sec:results}
We first applied Eq. \ref{eq:scalM} to derive masses of stars in our sample from \numax, \Dnu, and \Teff. As presented in the top row of Table \ref{tab:6791Mnocorr}, we find $\overline{M}_{\rm RGB}=1.23\pm0.02\, {\rm M}_\odot$ and $\overline{M}_{\rm RC }=1.03 \pm 0.03\,  {\rm M}_\odot$. The mass difference we obtain is thus $\Delta \overline{M}=0.19 \pm 0.04$ M$_\odot$.   
The value of $z=1.8-1.9$ reported in the first row of Table  \ref{tab:6791Mnocorr} suggests that  the scatter in the population is larger than expected from the formal uncertainties in the masses. This could indicate that uncertainties in the observed quantities are underestimated, provided, of course, that there is no significant intrinsic mass spread within the groups of RGB and RC stars.
The derived masses for RGB and RC stars are shown as a function of apparent $V$ magnitude in the top row of Fig. \ref{fig:masses6791}.

To check the robustness of the mass determination, we also estimated masses using Eq. \ref{eq:scalM2}, \ref{eq:scalM3} and \ref{eq:scalM4}, thereby including independent information on the distance to the cluster (see Sec. \ref{sec:data}). The results are presented in  Table \ref{tab:6791Mnocorr}, where we consider separately the contributions from random uncertainties and  (in parentheses) systematic errors from the distance.  We notice that the systematic uncertainty has little impact on $\Delta\overline{M}$ ($\leq 0.01\ {\rm M}_{\odot}$). The correlation between \teff\ and BC has been taken into account in the error estimates.

If we consider only RGB stars, excellent agreement is found between $\overline{M}_{\rm RGB}$ derived using different observables and scaling relations (see also Fig. \ref{fig:masses6791}, left panels).  
\begin{figure*}
\begin{center}
\begin{flushleft}
\vspace{0.5cm}
\end{flushleft}
\resizebox{.9\hsize}{!}{\includegraphics[]{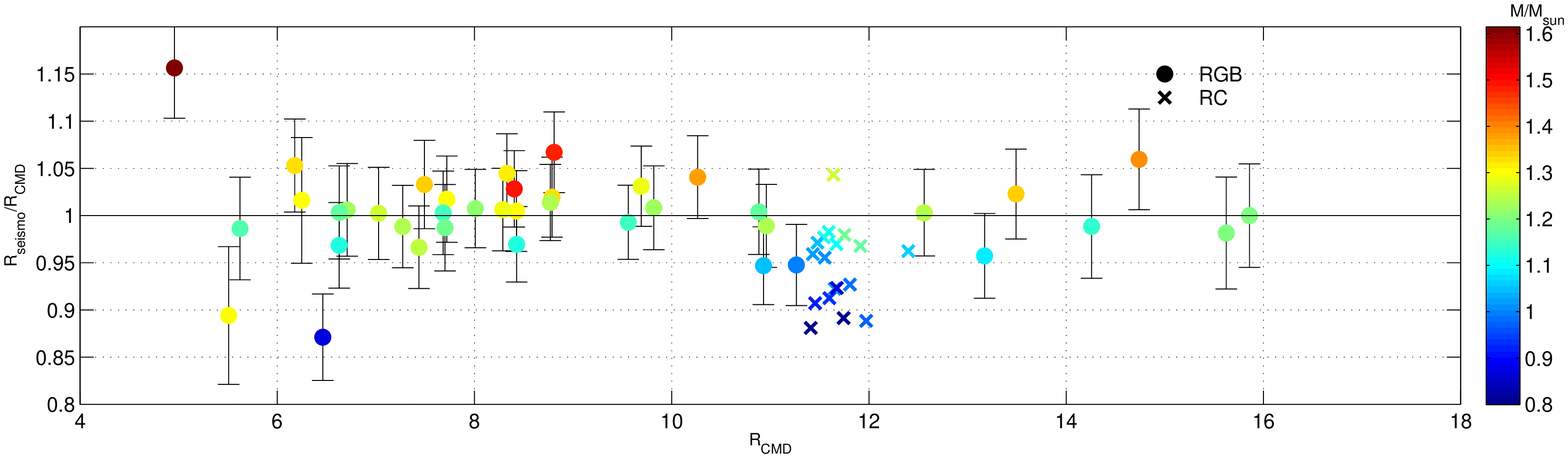}}
\caption{\small NGC6791: ratio between radii determined using $L$ and \Teff\ ($R_{\rm CMD}$), and those obtained via Eq. \ref{eq:scalR} ($R_{\rm seismo}$). The mass of each star determined via Eq. \ref{eq:scalM} is colour coded. }
\label{fig:radiusbefore}
\end{center}
\end{figure*}
\begin{figure*}
\begin{center}
\resizebox{.6\hsize}{!}{\includegraphics[]{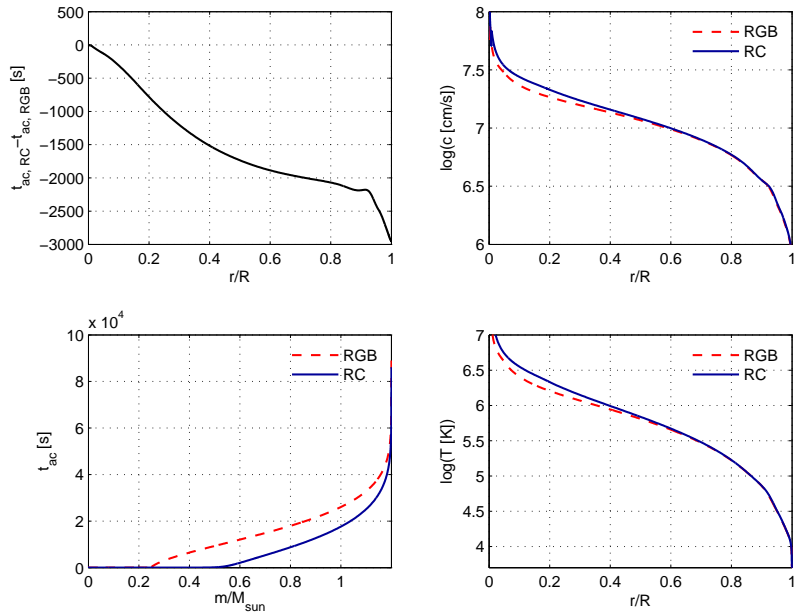}}
\caption{\small Upper-left panel: difference between the acoustic radius of RGB and RC models of same mass and radius as a function of relative radius. Right panels: sound speed and temperature stratification as a function of the normalised radius in the RGB model (red dashed line) and RC model (blue solid  line). Lower-left panel: acoustic radius as a function of mass (in solar units) for the RGB and RC models.}
\label{fig:acradius}
\end{center}
\end{figure*}
On the other hand, significant disagreement is found between $\overline{M}_{\rm RC}$ determined with different combinations of scaling relations.
As an additional test, and to investigate the reason for this discrepancy, we compared radii determined from $L$ and \Teff\ ($R_{\rm CMD}$), with those derived seismically through Eq. \ref{eq:scalR} ($R_{\rm seismo}$). As shown in Fig. \ref{fig:radiusbefore}, for RGB stars the two determinations of radii agree within 5\% (with an average relative difference $( R_{\rm CMD}-R_{\rm seismo})/R_{\rm CMD}=-0.15 \%$).  This comparison can also be carried out in terms of distance moduli, instead of radii. By coupling photometric constraints with $R_{\rm seismo}$ of the RGB stars, we find an average distance modulus $(m- M)_V=13.51 \pm 0.02$~mag, where the quoted uncertainty represents only the standard deviation from the mean. This determination is in excellent agreement with the value found by \citet{Brogaard2011} using eclipsing binaries ($(m- M)_V=13.51 \pm 0.06$, see Sec. \ref{sec:data}). The situation is different for RC stars, whose radii obtained from Eq. \ref{eq:scalR} are systematically smaller than those derived from the CMD,  with an average relative difference of  $-5\%$. This discrepancy could indicate  a systematic error in the scaling relations for RC stars.


\subsection{The $\Delta\nu$ scaling relation for RC and RGB stars}
\label{sec:correction}
Following the asymptotic approximation for acoustic oscillation modes, \Dnu\ is directly related to the sound-travel time in the stellar interior. We therefore expect \Dnu\ to depend on the stellar structure. Given that stars on the RGB have an internal temperature (hence sound speed) distribution significantly different from that of RC stars,  we investigated whether this difference could have an impact on the mass determination via the \Dnu\, scaling relation.

We used the code ATON \citep{Ventura2008} to compute stellar models representative of red giants in NGC6791, i.e. 1.2 M$_\odot$ models with initial heavy-element mass fraction $Z=0.04$, and initial helium mass fraction $Y=0.30$ \citep[see][for more details on the models]{Montalban2010}. The evolution of the models was followed through the helium flash. We then considered two models with the same radius  (within 0.1\%): one on the RGB and the other in the RC. We computed radial oscillation frequencies with the LOSC code \citep{Scuflaire2008b} and found that the RC model had a mean large frequency separation $\sim 3.3 \%$ larger than the RGB model, despite having the same mean density. 

The average large frequency separation is known to be directly related to the behaviour of the sound speed in the stellar interior, since $\Delta\nu\simeq1/(2\,T_{\rm ac})$ where $T_{\rm ac}$ is the total acoustic radius of the star. The latter is defined as  $T_{\rm ac}=t_{\rm ac}(R)$, with $t_{\rm ac}(r)=\int_0^r{{\rm d}r'/c(r')}$ being the acoustic radius, $c$ the sound speed, and $R$ the photospheric radius.

The upper-left panel of Fig \ref{fig:acradius} shows the layers contributing to the difference in the overall acoustic radius.  As shown in the upper-right panel, the sound speed in the RC model is on average higher (at a given fractional radius) than that of the RGB model, the main reason being the different temperature profile in the two models (lower-right panel).  The difference in acoustic radius becomes significant in regions as deep as $m/{\rm M}_\odot \simeq 0.25$, which is where the boundary of the helium core in the RGB model is located (lower-left panel). Finally, we note that while the largest contribution to the overall difference originates in the deep interior, near-surface regions ($r/R \gtrsim 0.9$) also contribute (by 0.8\%) to the total 3.5\% difference in $T_{\rm ac}$ (upper-left panel of Fig. \ref{fig:acradius}). 

These findings suggest that, if we intend to use the $\Delta\nu$ scaling to estimate stellar mean density, a relative correction has to be considered when dealing with RC and RGB stars. This relative correction is expected to be mass-dependent and to be larger for low-mass stars, which have significantly different internal structure when ascending the RGB compared to when they are in the core-He burning phase.  Deriving an accurate  theoretical correction of the \Dnu\ scaling, however, is beyond the scope of this paper, as additional models should be considered and detailed tests should be made of the internal structure of such stars. 
Guided by the representative modelling presented here, we instead adopted an empirical approach to correct the \Dnu\ scaling.

We determined the correction factor to the  large frequency separation by minimising the average relative difference $( R_{\rm CMD}-R_{\rm seismo})/R_{\rm CMD}$ for RC stars. The correction factor derived following this procedure is $2.7\%$, in good agreement with the correction suggested by the models. Correcting the \Dnu\ scaling for RC stars has a direct effect on $\overline{M}_{\rm RC}$ obtained with different scaling relations. After we applied the correction to the \Dnu\ scaling,  the revised average mass of RC stars computed with Eq. \ref{eq:scalM}, \ref{eq:scalM2}, and \ref{eq:scalM4} agreed within 1-$\sigma$ with the value obtained with Eq. \ref{eq:scalM3}, which does not contain \Dnu.

We note that the uncertainties in $R_{\rm seismo}$ and $R_{\rm CMD}$ imply an uncertainty of {1\%} in the correction factor to the \Dnu\ scaling relation. This leads to an additional source of systematic uncertainty in the mass estimate, which is largest (0.03 M$_{\odot}$) when using Eq. 3 to determine $\Delta\overline{M}$, and zero when using Eq. 6. Considering also potential differential temperature scale shifts between the RC and RGB (at the level of $1\%$), we adopted 0.04 M$_{\odot}$ as a conservative estimate of the additional systematic uncertainty in $\Delta\overline{M}$. 
Taking into account the uncertainties and scatter of the average mass of RC and RGB stars obtained with different scaling relations, we derived our best estimate of the difference between average masses of RGB and RC stars: $\Delta \overline{M}=0.09\pm 0.03$ (random) $\pm 0.04$ (systematic). 

Finally, to check for additional sources of bias due to the $T_{\rm eff}$ scale, we considered effective temperatures determined from $(V-I)$ colours. These $T_{\rm eff}$ introduce small but systematic trends in the derived quantities (e.g. mass vs. apparent magnitude) and worsen the agreement between masses and radii derived using different observational constraints. We find, nonetheless, that $\Delta \overline{M}$ obtained with the different temperatures scales agree within the quoted uncertainty range.

\section{NGC6819}
\label{sec:6819}
NGC6819 is an open cluster of near-solar metallicity \citep{Bragaglia2001}  estimated to be 2-2.5 Gyr old \citep{Kalirai2004, Basu2011}. Given that the cluster is significantly younger than NGC6791, a smaller RGB mass loss is expected. As for the case of NGC6791, we refer to \citet{Basu2011}, \citet{Hekker2011}, \citet{Stello2011b}, and \citet{Stello2011a} for a description of the asteroseismic data available for the stars in this cluster. 

We selected targets on the RGB and in the RC following the same criteria used in Sec. \ref{sec:data}, and retained in the final list 19 RGB stars and 13 RC stars. 
As for NGC6791, we adopted the effective temperatures determined using $(V-K)$ colours, the colour-temperature calibration of \citet{Ramirez2005}, and assuming a reddening $E(B-V)=0.15$~mag (see \citealt{Basu2011}) and [Fe/H]=0.

\begin{figure*}
\begin{center}
\resizebox{.85\hsize}{!}{\includegraphics[]{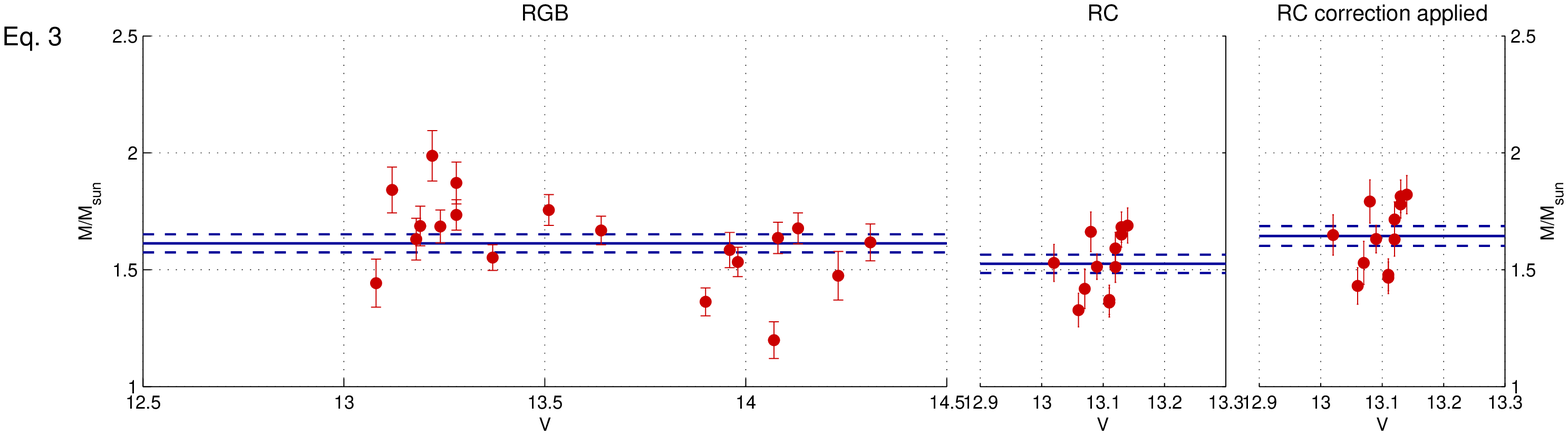}}
\resizebox{.85\hsize}{!}{\includegraphics[]{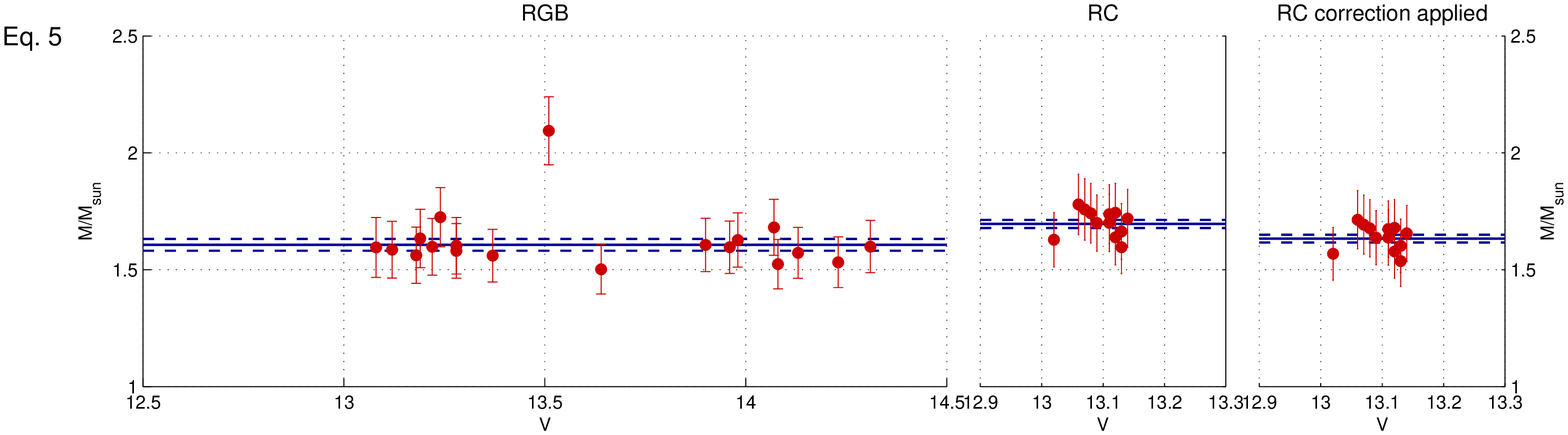}}
\resizebox{.85\hsize}{!}{\includegraphics[]{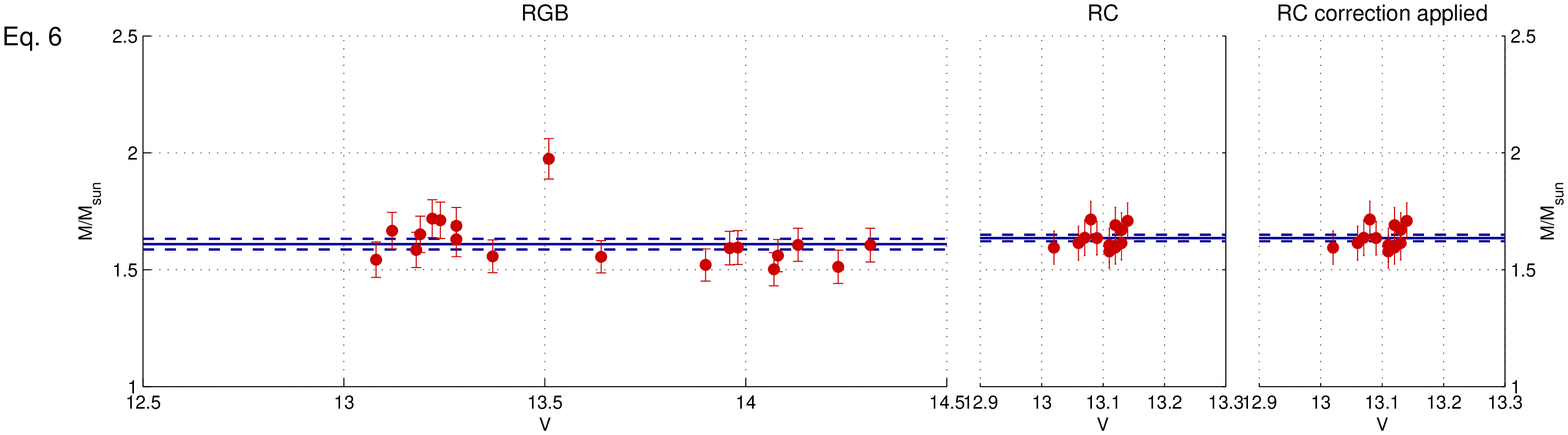}}
\resizebox{.85\hsize}{!}{\includegraphics[]{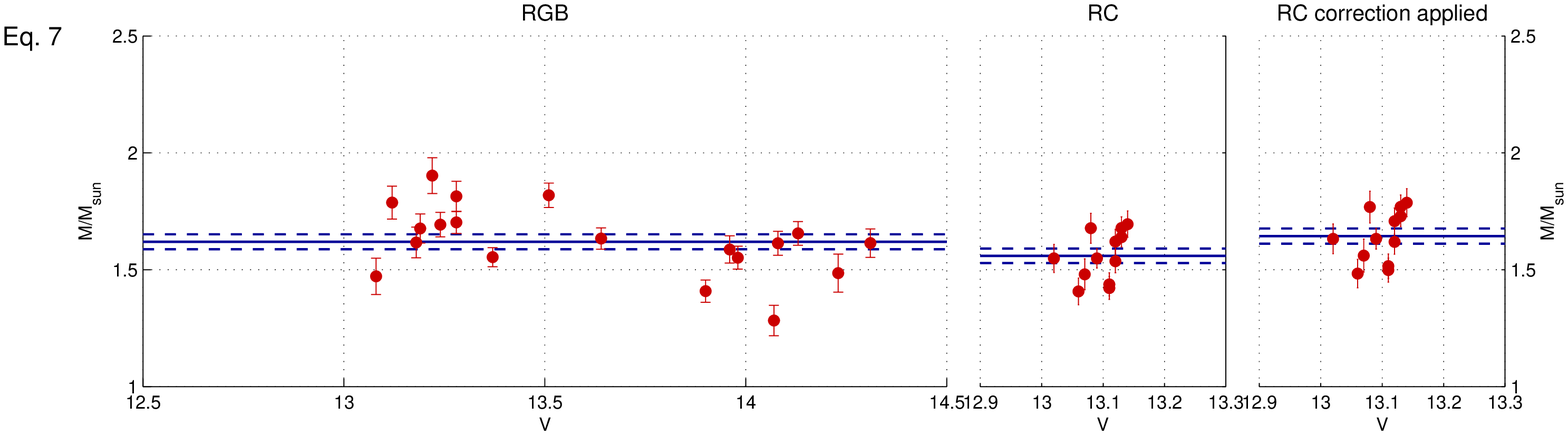}}

\vspace{0.5cm}
\caption{\small As Fig. \ref{fig:masses6791} but for red giants in NGC6819.}
\label{fig:masses6819}
\end{center}
\end{figure*}

We then performed a similar analysis as for NGC6791. However, since no accurate determination of the distance modulus was available for this cluster, we proceeded as in \citet{Basu2011} and estimated the distance modulus using the seismically determined radii of RGB stars and photometric constraints. We found $(m- M)_V=11.80 \pm 0.02$~mag, where the quoted uncertainty represents only the standard deviation from the mean (see \citealt{Basu2011} for a more detailed discussion of systematic uncertainties on the distance determination due to reddening).

We repeated the same steps as in Sec. \ref{sec:results} to determine the average mass of stars on the RGB and in the RC, and present our results in Table \ref{tab:compM6819nocorr} and Fig. \ref{fig:masses6819}.  We notice that, as in NGC6791,  $\overline{M}_{\rm RGB}$ is independent of the scaling used, but significant discrepancies appear in the estimated $\overline{M}_{\rm RC}$.  
Analogously to the case described in Sec \ref{sec:correction}, we used  $1.6\ {\rm M}_\odot$ stellar models and found that a relative correction to the \Dnu\ scaling was appropriate, and that the radii of RC stars are systematically smaller than those obtained using \Teff\ and $L$ (although we note that $L$ is not an independent quantity, as it is determined using Eq. \ref{eq:scalR}).   

We found that the correction factor for \Dnu\ that minimises the relative difference between $R_{\rm CMD}$ and $R_{\rm seismo}$ is 1.9\%, which is smaller than {for} NGC6791. Again, this is in qualitative agreement with the correction suggested by 1.6 M$_\odot$ models,  as the differences in the sound speed profile between RC and RGB stars of similar luminosity decreases when higher masses are considered. 

By combining results from different scaling relations, we derive $\Delta\overline{M}=-0.03\pm0.04 \ {\rm M}_\odot$ for NGC6819.  Due to the lack of an independent and accurate distance measurement, however, we consider this result less robust than the determination of $\Delta\overline M $ in NGC6791. 

\begin{table}
 \begin{tabular}{lcccccc}
\hline
Eq. &  $\overline{M}_{\rm RGB}\pm\sigma_{\overline{M}}$ & $z_{\rm RGB}$ & $\overline{M}_{\rm RC}\pm\sigma_{\overline{M}}$ & $z_{\rm RC}$ & $\Delta \overline{M}\pm\sigma_{\Delta\overline{M}}$  \\
\hline
 (\ref{eq:scalM}) &     $1.61 \pm     0.04$ &      2.3  &    $1.52 \pm     0.04 $   &  2.0  &    $0.09  \pm    0.06$\\ 
 (\ref{eq:scalM2}) &   $  1.61 \pm      0.03$&      0.9 &     $ 1.70\pm       0.02$  &    0.5 &  $   -0.09\pm       0.03$\\ 
 (\ref{eq:scalM3}) &   $  1.61  \pm     0.02$&      1.4  &   $ 1.64\pm       0.01$  &    0.6  &  $  -0.03\pm       0.03$ \\
 (\ref{eq:scalM4}) &   $  1.62  \pm     0.03$&      2.5  &    $1.56\pm       0.03 $ &    2.1  &   $ 0.06 \pm      0.04$ \\
\hline

 \end{tabular}
\caption{\small NGC6819: average mass of stars in the RGB and RC estimated using different observational constraints and scaling relations (Equations \ref{eq:scalM}, \ref{eq:scalM2},
\ref{eq:scalM3}, and \ref{eq:scalM4}). No correction to the  \Dnu\ scaling was applied to clump stars. }\label{tab:compM6819nocorr}

 \begin{tabular}{lccc}
\hline
Eq. &   $\overline{M}_{\rm RC}\pm\sigma_{\overline{M}}$ & $z_{\rm RC}$ & $\Delta \overline{M}\pm\sigma_{\Delta\overline{M}}$  \\
\hline
 (\ref{eq:scalM}) &     $1.65 \pm     0.04 $   &  2.0  &    $-0.03  \pm    0.06$\\ 
 (\ref{eq:scalM2}) &   $ 1.63\pm       0.02$  &    0.5 &  $   -0.03\pm       0.03$\\ 
 (\ref{eq:scalM3}) &     $ 1.64\pm       0.01$  &    0.7  &  $  -0.03\pm       0.03$ \\
 (\ref{eq:scalM4}) &   $1.64\pm       0.03 $ &    2.1  &   $ -0.02 \pm      0.05$ \\
\hline
 \end{tabular}
\caption{\small As in Table \ref{tab:compM6819nocorr},  after applying the 1.9\% suggested correction to the \Dnu\ scaling for clump stars.}\label{tab:compM6819corr}


\end{table}

\section{Inferences on the mass-loss rate}
\label{sec:massloss}
We now compare the difference between the average mass of low-luminosity RGB and  RC stars to theoretical predictions using a revised version of the  YZVAR\footnote{\texttt{http://stev.oapd.inaf.it/YZVAR/}} isochrones and stellar evolutionary sequences (\citealt{Bertelli2008};  Bertelli 2011, private communication). We varied the coefficient $\eta$ describing the efficiency of the RGB mass-loss rate, which was implemented in the models \citep[see][]{Bertelli1994} following  Reimers' (1975a) 
empirical formulation:
\begin{equation}
\label{eq:reimers}
\frac{{\rm d} M}{{\rm d} t}=1.27\,10^{-5}\,\eta\, M^{-1} L^{1.5} \,T_{\rm eff}^{-2}{\rm ,}
\end{equation}
where $M$ and $L$ are expressed in solar units, $T_{\rm eff}$ in K, and $t$ in yr.  A commonly adopted value of the parameter describing the mass-loss efficiency is $\eta=0.4$, see e.g. \citet{Renzini1988}.

We considered three scenarios: isochrones computed without mass loss ($\eta=0$), with moderate mass loss ($\eta=0.35$), and with an extreme value of the mass-loss efficiency ($\eta=0.7$). We notice that for these relatively young clusters, such a high mass-loss rate is not enough to suppress the RC and AGB phases, in contrast to old GCs \citep[see e.g.][]{Renzini1988}.

\begin{figure}
\begin{center}
\resizebox{1.\hsize}{!}{\includegraphics[]{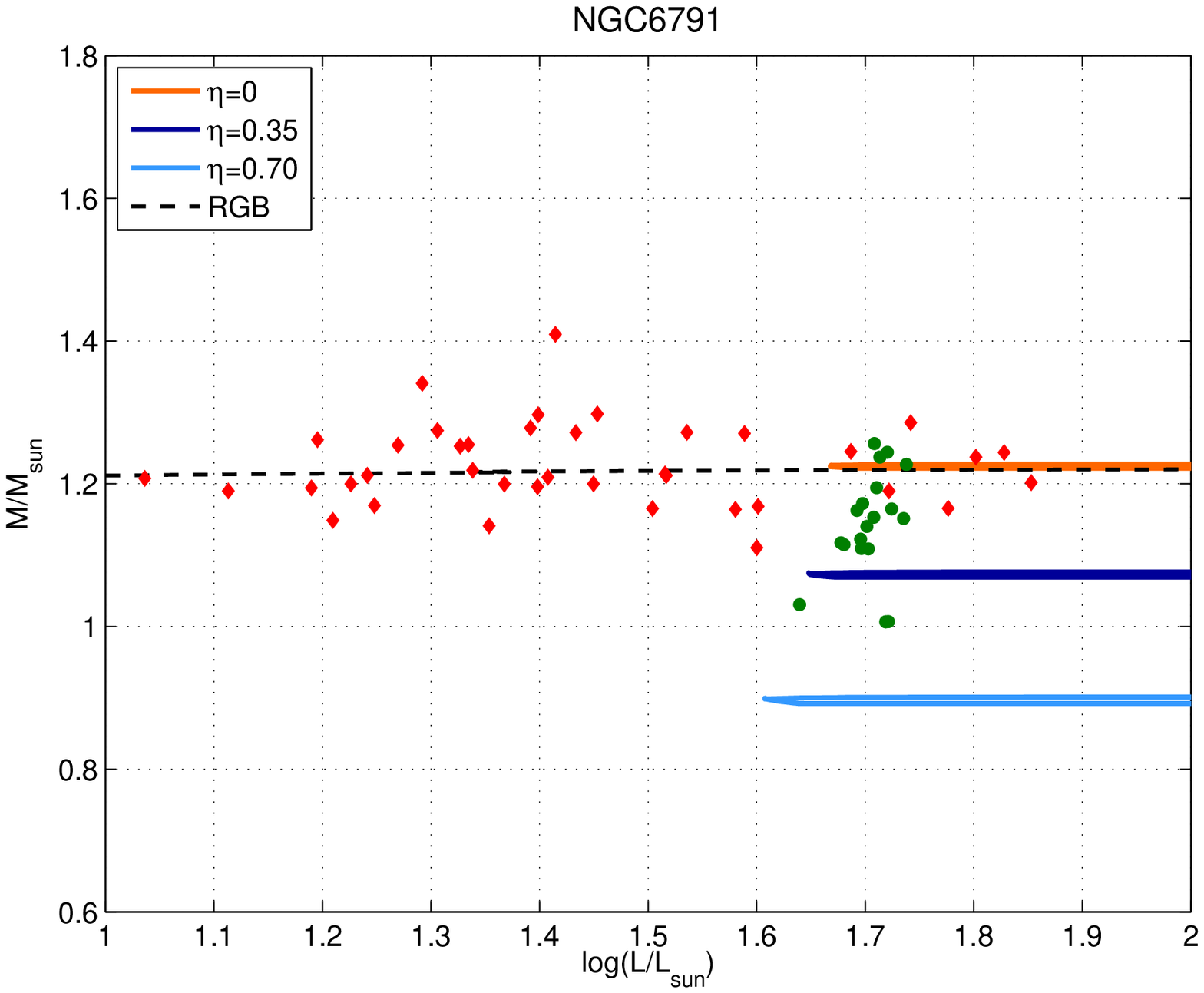}}
\resizebox{1.\hsize}{!}{\includegraphics[]{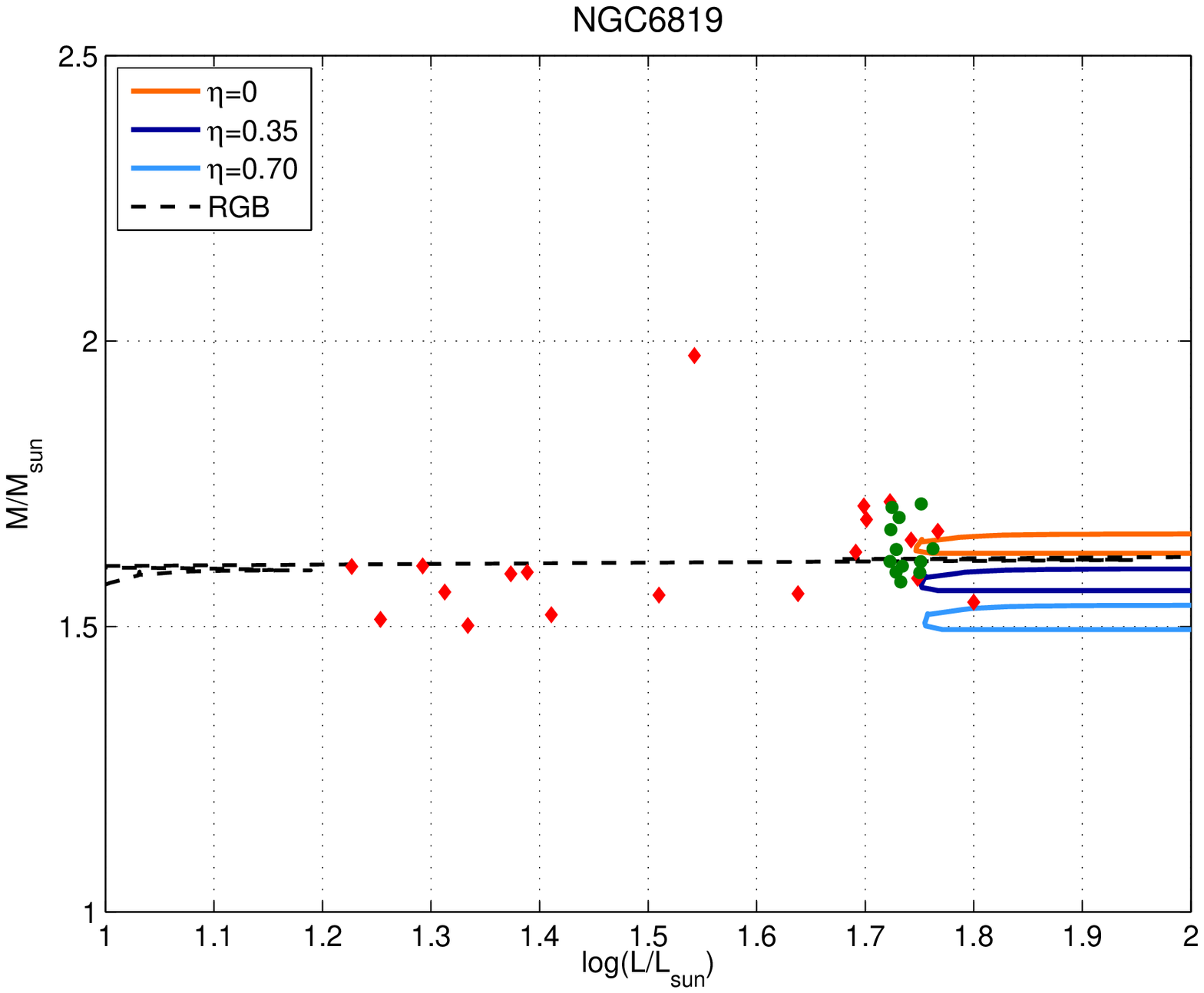}}

\caption{\small {Upper panel:} Masses of giants in NGC6791 derived from Eq. \ref{eq:scalM3} as a function of luminosity.  Red and green points indicate RGB and RC stars.  Lines represent the 7-Gyr isochrone computed with different mass-loss rates. The RGB part of the isochrone is indicated with a dashed line, while solid lines represent the evolution from  the He flash towards the zero-age HB, and onwards. {Lower panel:} Same as the upper panel, but for stars in NGC6819. For this cluster, a 2.1 Gyr, solar-metallicity isochrone was used.}
\label{fig:mloss6791}
\end{center}
\end{figure}

For NGC6791 we considered a 7-Gyr isochrone, computed with initial helium and heavy-elements mass fractions $Y=0.30$ and $Z=0.04$. In the case of NGC6819 we employed a 2.1-Gyr isochrone  with $Y=0.28$ and $Z=0.017$. 
Comparisons between observed and theoretical mass-luminosity diagrams are shown in Fig. \ref{fig:mloss6791}.
In these plots the masses of individual stars were determined by Eq. \ref{eq:scalM3}, and the luminosities were estimated by combining $T_{\rm eff}$ and the photospheric radius obtained via Eq. \ref{eq:scalR}. The Reimers mass-loss law predicts significant mass loss only near the RGB tip,  hence no appreciable effect is expected {($\Delta M \lesssim 0.01$ M$_\odot$)} in the RGB stars we considered, which have relatively low luminosities. With our present measurements we have no evidence for the time-dependence of the mass loss, as no significant mass gradient is found on the RGB of NGC6791. {Current uncertainties, however, do not allow us to constrain the mass loss  in this portion of the RGB to better than the total mass loss derived. }

\begin{figure}
\begin{center}
\resizebox{1.\hsize}{!}{\includegraphics[]{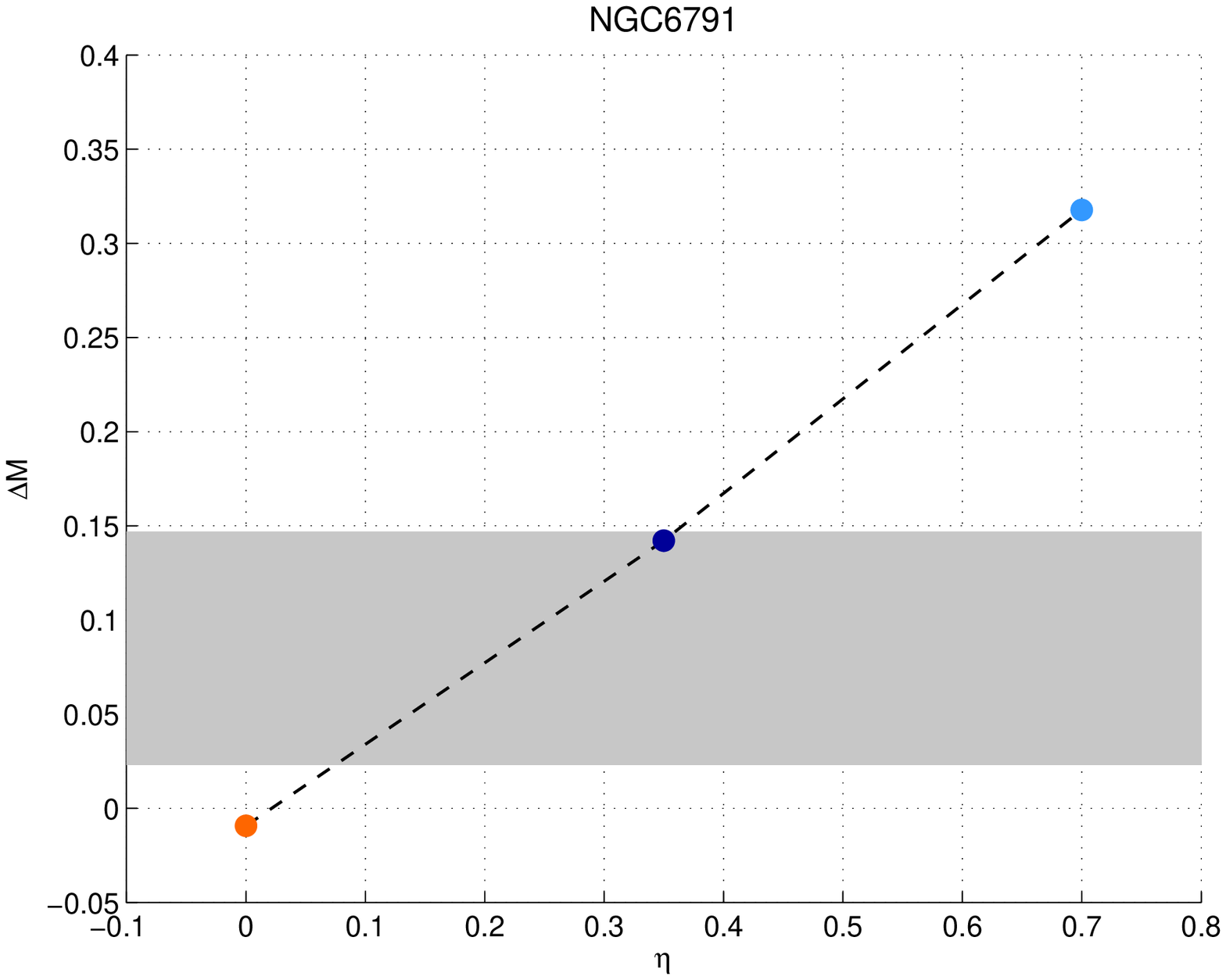}}
\resizebox{1.\hsize}{!}{\includegraphics[]{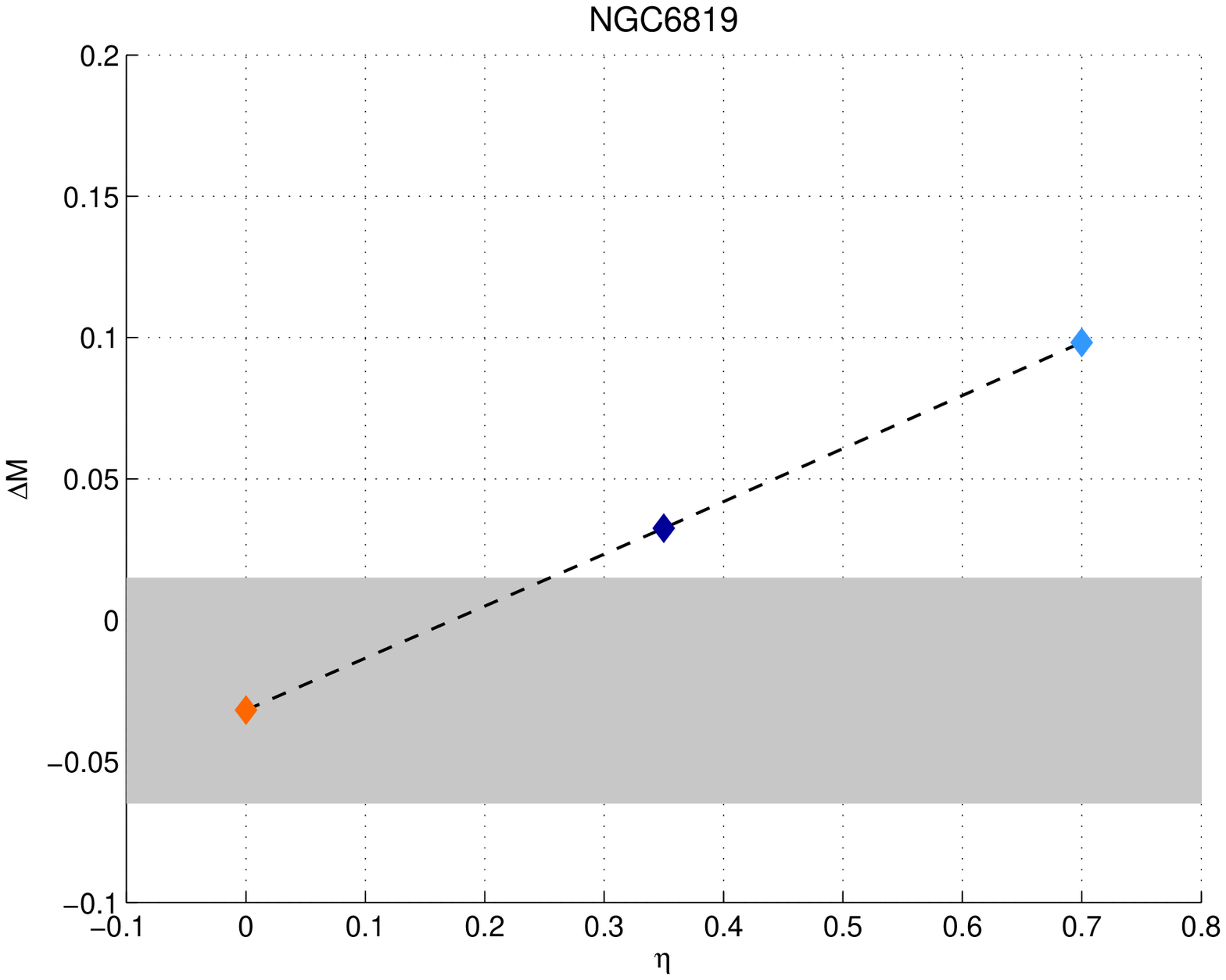}}

\caption{\small {Upper panel:} The dots connected by a dashed line shows the difference ($\Delta M$) between the average mass of RGB stars (in the luminosity domain of the targets) and RC stars as a function of the $\eta$ parameter, resulting from a  7 Gyr model isochrone. The grey areas represent the 1-$\sigma$  region of the observed $\Delta M$. {Lower panel:} Same plot as the upper panel, but for stars in NGC6819 and a 2.1 Gyr isochrone.}
\label{fig:mlossall}
\end{center}
\end{figure}

The comparison between observed and theoretically expected mass difference between RGB and RC stars is presented in Fig. \ref{fig:mlossall}. 
This first comparison shows that we can exclude very efficient RGB mass loss, and that the data are compatible with mass-loss rates described with a Reimers $\eta$ parameter smaller than $\sim$ 0.35 and, in the better constrained case of NGC6791, higher than 0.1. We note that the relation between $\Delta\overline M$ and $\eta$ predicted by the models has little dependence on the assumed age and chemical composition (e.g. in the case of NGC6791 we checked the robustness of our main conclusion considering also isochrones of 6 and 9 Gyr, $Y=0.28,0.32$ and $Z=0.025, 0.030$).

We finally note that in the case of NGC6819, a negative $\Delta\overline M$ is expected from the models with little or no mass loss.  For such a young cluster  $\Delta\overline M$ underestimates the integrated mass loss on the RGB, since the initial mass of stars in the RC is higher than that of stars on the RGB. {In the case of the 2.1-Gyr isochrone considered in Figs. \ref{fig:mloss6791} and \ref{fig:mlossall}, the difference in average progenitor mass between RGB  and RC stars is $\simeq-0.03$ M$_\odot$, hence the observed mass difference is compatible with no mass loss, as shown in the lower panel of Fig. \ref{fig:mlossall}.}


\section{Summary and prospects}
\label{sec:summary}
We used the average seismic parameters characterising solar-like oscillations (\numax\ and \Dnu) to estimate the masses of stars in the  RC and on the low-luminosity RGB ($L < L({\rm RC})$) of the open clusters NGC6791 and NGC6819. 

In NGC6791, \numax\ and \Dnu\ were coupled with additional constraints (photometric magnitudes, \Teff, and distance modulus from eclipsing binaries) to measure stellar radii, and to determine the average mass of stars on the RGB and in the RC. When considering RGB stars, we found good agreement between different determinations of radius and mass, while in the RC a larger scatter between $\overline M_{\rm RC}$ was found, together with a systematic underestimation of the radii obtained via Eq. \ref{eq:scalR} compared to those determined using photometric constraints and the independent distance determination obtained by \citet{Brogaard2011}. Supported by a comparison between RC and RGB models,  we ascribed this discrepancy to the \Dnu\ scaling (Eq. \ref{eq:dnu}), and argued that a relative correction of $\sim$ 3\% to \Dnu\ between RC and RGB stars should be adopted for stars in this cluster.

Taking into account the uncertainties in the observational constraints, we found that in NGC6791 the difference between the average mass of RGB and RC stars  is small but significant ($\Delta \overline{M}=0.09\pm 0.03$ (random) $\pm 0.04$ (systematic) M$_\odot$).  In the context of the controversial and highly debated results on the mass-loss rates in NGC6791 (see Sec. \ref{sec:mloss6791}), our direct estimate of $\Delta \overline{M}$  is incompatible with an extreme mass loss for this metal-rich cluster. If we describe the mass-loss rate with Reimers' prescription (see Eq. \ref{eq:reimers}), and assume a 7-Gyr isochrone, we find that the observed $\Delta\overline M$ is compatible with a mass-loss efficiency parameter of $0.1 \lesssim \eta \lesssim 0.3$. 
The low mass-loss of the RC stars, coupled with a RC mass-loss dispersion lower than our measurement errors, provides support for our assumption (see Sec. \ref{sec:data}) that either no helium-burning stars exist with masses significantly lower than the measured RC stars or, if they do, they must originate from a very different evolution.

Constraints on the RGB mass loss set by the analysis of NGC6819 are less compelling, largely because this cluster is too young to exhibit a significant mass loss (at least according to Reimers' formula). Moreover, no accurate and independent information on the distance was available for this cluster.

In the future, a reliable age determination of the clusters obtained by combining all available constraints will potentially allow detailed tests of  different formulations  of the RGB mass-loss rate \citep[e.g.][]{Catelan2000, Schroder2005, Schroder2007}, eventually leading to a better understanding of the physical mechanism responsible for the mass loss itself. Moreover, as the  duration of $Kepler$ observations increases, it will soon be possible to study in detail the pulsational properties of stars with luminosities higher than the RC. These targets could provide the link between the stars studied in this paper and  the luminous ($L\gtrsim10^3 {\rm L}_\odot$) RGB and AGB giants in which a connection between mass loss and pulsations has been suggested by several studies in the literature \citep[see e.g.,][and references therein]{Lebzelter2005}.

In this work we reached a point where possible systematic biases in the \Dnu\ and \numax\ scaling relations may become the largest source of uncertainty in the determination of stellar parameters. As shown in Sec. \ref{sec:correction}, we have found evidence for systematic differences in the \Dnu\ scaling relation between He-burning and H-shell burning giants. A more systematic study using a larger set of stellar models is needed to address this issue in detail.  {In particular, robust model-based corrections to the \Dnu\ scaling relation would allow testing in detail possible systematic errors in the \numax\ scaling, hence improving the accuracy of the mass determination.}

{The frequency of maximum power will in general depend on the dynamical properties of near-surface convective layers, and on the complex interaction between oscillation modes and convection. The published relations, however, assume a simple scaling of \numax\ with the acoustic cutoff frequency in the atmosphere, and simplifying assumptions about the latter.  It is therefore possible that evolutionary-state dependent corrections may be required for \numax\ also.  However, the close correspondence between the radius anomalies expected from \Dnu\ effects and those seen in NGC6791 suggests either that the impact of such corrections is modest, or that there is some effective cancellation of terms in the radius scalings.  The \numax\ scaling should be investigated, but a proper formulation would involve detailed convective modelling which is outside the scope of the current work.  One promising approach would be to search for residual differences in stars of known mass and radius once more robust theoretical corrections to \Dnu\ are applied, and there is work in preparation pursuing this avenue of approach. }

In this context, pulsating red giants with independent masses and/or radii (e.g. binary members and nearby giants) would allow   calibration of the  \Dnu\ and \numax\ scaling relations, and provide an unprecedented, and potentially very accurate, means of determining stellar parameters.

Additional observational constraints that will improve the precision and accuracy of our conclusions include {\it Kepler} data that will soon be available for most stars in the clusters. Moreover, the availability in the near future of not only average seismic parameters such as \numax\ and \Dnu, but also of accurate frequencies of individual oscillation modes in red giants, will allow more detailed tests of the scaling relations as well as of the internal structure of these stars.
Finally, spectroscopic constraints on \Teff\ will also be of great relevance to check the validity of the adopted \Teff\ scale, and to further test the robustness of the agreement  we find between global parameters of stars  determined using different constraints (asteroseismology, photometry, distance determined using eclipsing binaries).

\section*{Acknowledgements}
The authors acknowledge the {\it Kepler} Science Team and all those who have contributed to making the {\it Kepler} mission possible. Funding for the {\it Kepler} Discovery mission is provided by NASA's Science Mission Directorate. 
AM acknowledges the support of the School of Physics and Astronomy, University of Birmingham.  KB acknowledges financial support from the Carlsberg Foundation. DS acknowledges support from the Australian Research Council. WJC and YE acknowledge support from the UK STFC.
AMS is supported by the  European  Union International Reintegration Grant  PIRG-GA-2009-247732, the MICINN  grant AYA08-1839/ESP, by
the  ESF EUROCORES Program  EuroGENESIS (MICINN  grant EUI2009-04170),  by SGR grants  of   the  Generalitat  de   Catalunya  and  by  the   EU-FEDER  funds.
SH acknowledges financial support from the Netherlands Organisation for Scientific Research (NWO). 
This project has been supported by the `Lend\"ulet' program
of the Hungarian Academy of Sciences and the Hungarian OTKA
grants K83790 and MB08C 81013. R.Sz. thanks the support of
the J\'anos Bolyai Research Scholarship.
The research leading to these results has received funding 
from the European Community's Seventh Framework Programme
(FP7/2007-2013) under grant agreement no. 269194.
\bibliographystyle{mn2e_new}
\small
\bibliography{andrea_m}

\begin{thebibliography}{88}
\expandafter\ifx\csname natexlab\endcsname\relax\def\natexlab#1{#1}\fi

\bibitem[{Althaus {et~al}\mbox{.}(2010)Althaus, Garc\'{\i}a-Berro, Renedo,
  Isern, C\'{o}rsico, \& Rohrmann}]{althaus2010}
Althaus L.~G., Garc\'{\i}a-Berro E., Renedo I., Isern J., C\'{o}rsico A.~H.,
  Rohrmann R.~D., 2010, \apj, 719, 612

\bibitem[{{Basu} {et~al}\mbox{.}(2011){Basu}, {Grundahl}, {Stello},
  {Kallinger}, {Hekker}, {Mosser}, {Garc{\'{\i}}a}, {Mathur}, {Brogaard},
  {Bruntt}, {Chaplin}, {Gai}, {Elsworth}, {Esch}, {Ballot}, {Bedding},
  {Gruberbauer}, {Huber}, {Miglio}, {Yildiz}, {Kjeldsen},
  {Christensen-Dalsgaard}, {Gilliland}, {Fanelli}, {Ibrahim}, \&
  {Smith}}]{Basu2011}
{Basu} S. {et~al.}, 2011, \apjl, 729, L10

\bibitem[{{Bedding} {et~al}\mbox{.}(2011){Bedding}, {Mosser}, {Huber},
  {Montalb{\'a}n}, {Beck}, {Christensen-Dalsgaard}, {Elsworth},
  {Garc{\'{\i}}a}, {Miglio}, {Stello}, {White}, {De Ridder}, {Hekker}, {Aerts},
  {Barban}, {Belkacem}, {Broomhall}, {Brown}, {Buzasi}, {Carrier}, {Chaplin},
  {di Mauro}, {Dupret}, {Frandsen}, {Gilliland}, {Goupil}, {Jenkins},
  {Kallinger}, {Kawaler}, {Kjeldsen}, {Mathur}, {Noels}, {Aguirre}, \&
  {Ventura}}]{Bedding2011}
{Bedding} T.~R. {et~al.}, 2011, \nat, 471, 608

\bibitem[{Bedin {et~al}\mbox{.}(2008{\natexlab{a}})Bedin, King, Anderson,
  Piotto, Salaris, Cassisi, \& Serenelli}]{Bedin2008}
Bedin L.~R., King I.~R., Anderson J., Piotto G., Salaris M., Cassisi S.,
  Serenelli A., 2008{\natexlab{a}}, \apj, 678, 1279

\bibitem[{Bedin {et~al}\mbox{.}(2008{\natexlab{b}})Bedin, Salaris, Piotto,
  Cassisi, Milone, Anderson, \& King}]{bedin2008a}
Bedin L.~R., Salaris M., Piotto G., Cassisi S., Milone A.~P., Anderson J., King
  I.~R., 2008{\natexlab{b}}, \apj, 679, L29

\bibitem[{Bedin {et~al}\mbox{.}(2005)Bedin, Salaris, Piotto, King, Anderson,
  Cassisi, \& Momany}]{bedin2005}
Bedin L.~R., Salaris M., Piotto G., King I.~R., Anderson J., Cassisi S., Momany
  Y., 2005, \apj, 624, L45

\bibitem[{{Belkacem} {et~al}\mbox{.}(2011){Belkacem}, {Goupil}, {Dupret},
  {Samadi}, {Baudin}, {Noels}, \& {Mosser}}]{Belkacem2011}
{Belkacem} K., {Goupil} M.~J., {Dupret} M.~A., {Samadi} R., {Baudin} F.,
  {Noels} A., {Mosser} B., 2011, \aap, 530, A142

\bibitem[{{Bertelli} {et~al}\mbox{.}(1994){Bertelli}, {Bressan}, {Chiosi},
  {Fagotto}, \& {Nasi}}]{Bertelli1994}
{Bertelli} G., {Bressan} A., {Chiosi} C., {Fagotto} F., {Nasi} E., 1994, \aaps,
  106, 275

\bibitem[{{Bertelli} {et~al}\mbox{.}(2008){Bertelli}, {Girardi}, {Marigo}, \&
  {Nasi}}]{Bertelli2008}
{Bertelli} G., {Girardi} L., {Marigo} P., {Nasi} E., 2008, \aap, 484, 815

\bibitem[{Bowen(1988)}]{Bowen1988}
Bowen G.~H., 1988, \apj, 329, 299

\bibitem[{{Boyer} {et~al}\mbox{.}(2009){Boyer}, {McDonald}, {van Loon},
  {Gordon}, {Babler}, {Block}, {Bracker}, {Engelbracht}, {Hora}, {Indebetouw},
  {Meade}, {Meixner}, {Misselt}, {Oliveira}, {Sewilo}, {Shiao}, \&
  {Whitney}}]{Boyer2009}
{Boyer} M.~L. {et~al.}, 2009, \apj, 705, 746

\bibitem[{Boyer {et~al}\mbox{.}(2010)Boyer, van Loon, McDonald, Gordon, Babler,
  Block, Bracker, Engelbracht, Hora, Indebetouw, Meade, Meixner, Misselt,
  Sewilo, Shiao, \& Whitney}]{Boyer2010}
Boyer M.~L. {et~al.}, 2010, \apj, 711, L99

\bibitem[{{Bragaglia} {et~al}\mbox{.}(2001){Bragaglia}, {Carretta}, {Gratton},
  {Tosi}, {Bonanno}, {Bruno}, {Cal{\`i}}, {Claudi}, {Cosentino}, {Desidera},
  {Farisato}, {Rebeschini}, \& {Scuderi}}]{Bragaglia2001}
{Bragaglia} A. {et~al.}, 2001, \aj, 121, 327

\bibitem[{{Brogaard} {et~al}\mbox{.}(2011){Brogaard}, {Bruntt}, {Grundahl},
  {Clausen}, {Frandsen}, {Vandenberg}, \& {Bedin}}]{Brogaard2011}
{Brogaard} K., {Bruntt} H., {Grundahl} F., {Clausen} J.~V., {Frandsen} S.,
  {Vandenberg} D.~A., {Bedin} L.~R., 2011, \aap, 525, A2

\bibitem[{Brown {et~al}\mbox{.}(1991)Brown, Gilliland, Noyes, \&
  Ramsey}]{Brown1991}
Brown T.~M., Gilliland R.~L., Noyes R.~W., Ramsey L.~W., 1991, \apj, 368, 599

\bibitem[{Brown {et~al}\mbox{.}(2001)Brown, Sweigart, Lanz, Landsman, \&
  Hubeny}]{brown2001}
Brown T.~M., Sweigart A.~V., Lanz T., Landsman W.~B., Hubeny I., 2001, \apj,
  562, 368

\bibitem[{Carretta {et~al}\mbox{.}(2009)Carretta, Bragaglia, Gratton, \&
  Lucatello}]{Carretta2009}
Carretta E., Bragaglia A., Gratton R., Lucatello S., 2009, \aap, 505, 139

\bibitem[{Castellani \& Castellani(1993)}]{castellani1993}
Castellani M., Castellani V., 1993, \apj, 407, 649

\bibitem[{Catelan(2000)}]{Catelan2000}
Catelan M., 2000, \apj, 531, 826

\bibitem[{Catelan(2009)}]{Catelan09}
---, 2009, \apss, 320, 261

\bibitem[{Chaboyer, Green \& Liebert(1999)Chaboyer, Green, \&
  Liebert}]{Chaboyer1999}
Chaboyer B., Green E.~M., Liebert J., 1999, The Astronomical Journal, 117, 1360

\bibitem[{{Chaplin} {et~al}\mbox{.}(1998){Chaplin}, {Elsworth}, {Isaak},
  {Lines}, {McLeod}, {Miller}, \& {New}}]{Chaplin1998}
{Chaplin} W.~J., {Elsworth} Y., {Isaak} G.~R., {Lines} R., {McLeod} C.~P.,
  {Miller} B.~A., {New} R., 1998, \mnras, 300, 1077

\bibitem[{{D'Antona} \& {Caloi}(2004)}]{D'Antona2004}
{D'Antona} F., {Caloi} V., 2004, \apj, 611, 871

\bibitem[{D'Antona {et~al}\mbox{.}(2002)D'Antona, Caloi, Montalb\'an, Ventura,
  \& Gratton}]{D'Antona2002}
D'Antona F., Caloi V., Montalb\'an J., Ventura P., Gratton R., 2002, \aap, 395,
  69

\bibitem[{Deloye \& Bildsten(2002)}]{deloye2002}
Deloye C.~J., Bildsten L., 2002, \apj, 580, 1077

\bibitem[{{Flower}(1996)}]{Flower1996}
{Flower} P.~J., 1996, \apj, 469, 355

\bibitem[{Friel \& Janes(1993)}]{Friel1993}
Friel E.~D., Janes K.~A., 1993, \aap, 267, 75

\bibitem[{{Gai} {et~al}\mbox{.}(2011){Gai}, {Basu}, {Chaplin}, \&
  {Elsworth}}]{Gai2011}
{Gai} N., {Basu} S., {Chaplin} W.~J., {Elsworth} Y., 2011, \apj, 730, 63

\bibitem[{{Garc{\'{\i}}a} {et~al}\mbox{.}(2011){Garc{\'{\i}}a}, {Hekker},
  {Stello}, {Guti{\'e}rrez-Soto}, {Handberg}, {Huber}, {Karoff},
  {Uytterhoeven}, {Appourchaux}, {Chaplin}, {Elsworth}, {Mathur}, {Ballot},
  {Christensen-Dalsgaard}, {Gilliland}, {Houdek}, {Jenkins}, {Kjeldsen},
  {McCauliff}, {Metcalfe}, {Middour}, {Molenda-Zakowicz}, {Monteiro}, {Smith},
  \& {Thompson}}]{Garcia2011}
{Garc{\'{\i}}a} R.~A. {et~al.}, 2011, \mnras, L239

\bibitem[{Garc\'{\i}a-Berro {et~al}\mbox{.}(2010)Garc\'{\i}a-Berro, Torres,
  Althaus, Renedo, Lor\'{e}n-Aguilar, C\'{o}rsico, Rohrmann, Salaris, \&
  Isern}]{garcia-berro2010}
Garc\'{\i}a-Berro E. {et~al.}, 2010, Nature, 465, 194

\bibitem[{{Garc{\'{\i}}a-Berro} {et~al}\mbox{.}(2011){Garc{\'{\i}}a-Berro},
  {Torres}, {Renedo}, {Camacho}, {Althaus}, {C{\'o}rsico}, {Salaris}, \&
  {Isern}}]{garcia-berro2011}
{Garc{\'{\i}}a-Berro} E., {Torres} S., {Renedo} I., {Camacho} J., {Althaus}
  L.~G., {C{\'o}rsico} A.~H., {Salaris} M., {Isern} J., 2011, ArXiv e-prints

\bibitem[{{Goldberg, L.}(1979)}]{Goldberg1979}
{Goldberg, L.}, 1979, Royal Astronomical Society, 20, 361

\bibitem[{Gratton {et~al}\mbox{.}(2001)Gratton, Bonifacio, Bragaglia, Carretta,
  Castellani, Centurion, Chieffi, Claudi, Clementini, D'Antona, Desidera,
  Fran\c{c}ois, Grundahl, Lucatello, Molaro, Pasquini, Sneden, Spite, \&
  Straniero}]{Gratton2001}
Gratton R.~G. {et~al.}, 2001, \aap, 369, 87

\bibitem[{Han {et~al}\mbox{.}(2003)Han, Podsiadlowski, Maxted, \&
  Marsh}]{han2003}
Han Z., Podsiadlowski P., Maxted P. F.~L., Marsh T.~R., 2003, Monthly Notices
  of the Royal Astronomical Society, 341, 669

\bibitem[{Hansen(2005)}]{hansen2005}
Hansen B. M.~S., 2005, \apj, 635, 522

\bibitem[{Hansen {et~al}\mbox{.}(2007)Hansen, Anderson, Brewer, Dotter,
  Fahlman, Hurley, Kalirai, King, Reitzel, Richer, Rich, Shara, \&
  Stetson}]{hansen2007}
Hansen B. M.~S. {et~al.}, 2007, \apj, 671, 380

\bibitem[{Harper(1996)}]{Harper1996}
Harper G., 1996, Cool stars; stellar systems; and the sun : 9 : Astronomical
  Society of the Pacific Conference Series, 109

\bibitem[{{Hekker} {et~al}\mbox{.}(2011){Hekker}, {Basu}, {Stello},
  {Kallinger}, {Grundahl}, {Mathur}, {Garc{\'{\i}}a}, {Mosser}, {Huber},
  {Bedding}, {Szab{\'o}}, {De Ridder}, {Chaplin}, {Elsworth}, {Hale},
  {Christensen-Dalsgaard}, {Gilliland}, {Still}, {McCauliff}, \&
  {Quintana}}]{Hekker2011}
{Hekker} S. {et~al.}, 2011, \aap, 530, A100

\bibitem[{{H{\"o}fner}(2009)}]{Hoefner2009}
{H{\"o}fner} S., 2009, in Astronomical Society of the Pacific Conference
  Series, Vol. 414, Cosmic Dust - Near and Far, {T.~Henning, E.~Gr{\"u}n, \&
  J.~Steinacker}, ed., p.~3

\bibitem[{{Hole} {et~al}\mbox{.}(2009){Hole}, {Geller}, {Mathieu}, {Platais},
  {Meibom}, \& {Latham}}]{Hole2009}
{Hole} K.~T., {Geller} A.~M., {Mathieu} R.~D., {Platais} I., {Meibom} S.,
  {Latham} D.~W., 2009, \aj, 138, 159

\bibitem[{{Huber} {et~al}\mbox{.}(2009){Huber}, {Stello}, {Bedding}, {Chaplin},
  {Arentoft}, {Quirion}, \& {Kjeldsen}}]{Huber2009}
{Huber} D., {Stello} D., {Bedding} T.~R., {Chaplin} W.~J., {Arentoft} T.,
  {Quirion} P.-O., {Kjeldsen} H., 2009, Communications in Asteroseismology,
  160, 74

\bibitem[{{Jenkins} {et~al}\mbox{.}(2010{\natexlab{a}}){Jenkins}, {Caldwell},
  {Chandrasekaran}, {Twicken}, {Bryson}, {Quintana}, {Clarke}, {Li}, {Allen},
  {Tenenbaum}, {Wu}, {Klaus}, {Middour}, {Cote}, {McCauliff}, {Girouard},
  {Gunter}, {Wohler}, {Sommers}, {Hall}, {Uddin}, {Wu}, {Bhavsar}, {Van Cleve},
  {Pletcher}, {Dotson}, {Haas}, {Gilliland}, {Koch}, \&
  {Borucki}}]{Jenkins2010b}
{Jenkins} J.~M. {et~al.}, 2010{\natexlab{a}}, \apjl, 713, L87

\bibitem[{{Jenkins} {et~al}\mbox{.}(2010{\natexlab{b}}){Jenkins}, {Caldwell},
  {Chandrasekaran}, {Twicken}, {Bryson}, {Quintana}, {Clarke}, {Li}, {Allen},
  {Tenenbaum}, {Wu}, {Klaus}, {Van Cleve}, {Dotson}, {Haas}, {Gilliland},
  {Koch}, \& {Borucki}}]{Jenkins2010a}
---, 2010{\natexlab{b}}, \apjl, 713, L120

\bibitem[{Judge \& Stencel(1991)}]{Judge1991}
Judge P.~G., Stencel R.~E., 1991, \apj, 371, 357

\bibitem[{Kalirai {et~al}\mbox{.}(2007)Kalirai, Bergeron, Hansen, Kelson,
  Reitzel, Rich, \& Richer}]{kalirai2007}
Kalirai J.~S., Bergeron P., Hansen B. M.~S., Kelson D.~D., Reitzel D.~B., Rich
  R.~M., Richer H.~B., 2007, \apj, 671, 748

\bibitem[{{Kalirai} {et~al}\mbox{.}(2009){Kalirai}, {Saul Davis}, {Richer},
  {Bergeron}, {Catelan}, {Hansen}, \& {Rich}}]{Kalirai2009}
{Kalirai} J.~S., {Saul Davis} D., {Richer} H.~B., {Bergeron} P., {Catelan} M.,
  {Hansen} B.~M.~S., {Rich} R.~M., 2009, \apj, 705, 408

\bibitem[{{Kalirai} \& {Tosi}(2004)}]{Kalirai2004}
{Kalirai} J.~S., {Tosi} M., 2004, \mnras, 351, 649

\bibitem[{{Kallinger} {et~al}\mbox{.}(2010){Kallinger}, {Weiss}, {Barban},
  {Baudin}, {Cameron}, {Carrier}, {De Ridder}, {Goupil}, {Gruberbauer},
  {Hatzes}, {Hekker}, {Samadi}, \& {Deleuil}}]{Kallinger2010}
{Kallinger} T. {et~al.}, 2010, \aap, 509, A77

\bibitem[{King {et~al}\mbox{.}(2005)King, Bedin, Piotto, Cassisi, \&
  Anderson}]{king2005}
King I.~R., Bedin L.~R., Piotto G., Cassisi S., Anderson J., 2005, The
  Astronomical Journal, 130, 626

\bibitem[{Kjeldsen \& Bedding(1995)}]{Kjeldsen1995}
Kjeldsen H., Bedding T.~R., 1995, \aap, 293, 87

\bibitem[{{Koch} {et~al}\mbox{.}(2010){Koch}, {Borucki}, {Basri}, {Batalha},
  {Brown}, {Caldwell}, {Christensen-Dalsgaard}, {Cochran}, {DeVore}, {Dunham},
  {Gautier}, {Geary}, {Gilliland}, {Gould}, {Jenkins}, {Kondo}, {Latham},
  {Lissauer}, {Marcy}, {Monet}, {Sasselov}, {Boss}, {Brownlee}, {Caldwell},
  {Dupree}, {Howell}, {Kjeldsen}, {Meibom}, {Morrison}, {Owen}, {Reitsema},
  {Tarter}, {Bryson}, {Dotson}, {Gazis}, {Haas}, {Kolodziejczak}, {Rowe}, {Van
  Cleve}, {Allen}, {Chandrasekaran}, {Clarke}, {Li}, {Quintana}, {Tenenbaum},
  {Twicken}, \& {Wu}}]{Koch2010}
{Koch} D.~G. {et~al.}, 2010, \apjl, 713, L79

\bibitem[{Lafon \& Berruyer(1991)}]{Lafon1991}
Lafon J. P.~J., Berruyer N., 1991, The \aap Review, 2, 249

\bibitem[{{Lebzelter} \& {Wood}(2005)}]{Lebzelter2005}
{Lebzelter} T., {Wood} P.~R., 2005, \aap, 441, 1117

\bibitem[{{McDonald} {et~al}\mbox{.}(2011{\natexlab{a}}){McDonald}, {Boyer},
  {van Loon}, \& {Zijlstra}}]{McDonald2011a}
{McDonald} I., {Boyer} M.~L., {van Loon} J.~T., {Zijlstra} A.~A.,
  2011{\natexlab{a}}, \apj, 730, 71

\bibitem[{{McDonald} {et~al}\mbox{.}(2011{\natexlab{b}}){McDonald}, {Boyer},
  {van Loon}, {Zijlstra}, {Hora}, {Babler}, {Block}, {Gordon}, {Meade},
  {Meixner}, {Misselt}, {Robitaille}, {Sewi{\l}o}, {Shiao}, \&
  {Whitney}}]{McDonald2011b}
{McDonald} I. {et~al.}, 2011{\natexlab{b}}, \apjs, 193, 23

\bibitem[{{McDonald} {et~al}\mbox{.}(2009){McDonald}, {van Loon}, {Decin},
  {Boyer}, {Dupree}, {Evans}, {Gehrz}, \& {Woodward}}]{McDonald2009}
{McDonald} I., {van Loon} J.~T., {Decin} L., {Boyer} M.~L., {Dupree} A.~K.,
  {Evans} A., {Gehrz} R.~D., {Woodward} C.~E., 2009, \mnras, 394, 831

\bibitem[{{McDonald} {et~al}\mbox{.}(2011{\natexlab{c}}){McDonald}, {van Loon},
  {Sloan}, {Dupree}, {Zijlstra}, {Boyer}, {Gehrz}, {Evans}, {Woodward}, \&
  {Johnson}}]{McDonald2011c}
{McDonald} I. {et~al.}, 2011{\natexlab{c}}, ArXiv e-prints

\bibitem[{{Montalb{\'a}n} {et~al}\mbox{.}(2010){Montalb{\'a}n}, {Miglio},
  {Noels}, {Scuflaire}, \& {Ventura}}]{Montalban2010}
{Montalb{\'a}n} J., {Miglio} A., {Noels} A., {Scuflaire} R., {Ventura} P.,
  2010, \apjl, 721, L182

\bibitem[{{Mosser} {et~al}\mbox{.}(2011){Mosser}, {Barban}, {Montalb{\'a}n},
  {Beck}, {Miglio}, {Belkacem}, {Goupil}, {Hekker}, {De Ridder}, {Dupret},
  {Elsworth}, {Noels}, {Baudin}, {Michel}, {Samadi}, {Auvergne}, {Baglin}, \&
  {Catala}}]{Mosser2011}
{Mosser} B. {et~al.}, 2011, \aap, 532, A86

\bibitem[{{Mosser} {et~al}\mbox{.}(2010){Mosser}, {Belkacem}, {Goupil},
  {Miglio}, {Morel}, {Barban}, {Baudin}, {Hekker}, {Samadi}, {De Ridder},
  {Weiss}, {Auvergne}, \& {Baglin}}]{Mosser2010}
---, 2010, \aap, 517, A22

\bibitem[{Mullan(1978)}]{Mullan1978}
Mullan D.~J., 1978, \apj, 226, 151

\bibitem[{Origlia {et~al}\mbox{.}(2002)Origlia, Ferraro, {Fusi Pecci}, \&
  Rood}]{Origlia2002}
Origlia L., Ferraro F.~R., {Fusi Pecci} F., Rood R.~T., 2002, \apj, 571, 458

\bibitem[{Origlia {et~al}\mbox{.}(2007)Origlia, Rood, Fabbri, Ferraro, {Fusi
  Pecci}, \& Rich}]{origlia2007}
Origlia L., Rood R.~T., Fabbri S., Ferraro F.~R., {Fusi Pecci} F., Rich R.~M.,
  2007, \apj, 667, L85

\bibitem[{Origlia {et~al}\mbox{.}(2010)Origlia, Rood, Fabbri, Ferraro, {Fusi
  Pecci}, Rich, \& Dalessandro}]{origlia2010}
Origlia L., Rood R.~T., Fabbri S., Ferraro F.~R., {Fusi Pecci} F., Rich R.~M.,
  Dalessandro E., 2010, \apj, 718, 522

\bibitem[{Origlia {et~al}\mbox{.}(2006)Origlia, Valenti, Rich, \&
  Ferraro}]{Origlia2006}
Origlia L., Valenti E., Rich R.~M., Ferraro F.~R., 2006, \apj, 646, 499

\bibitem[{{Peterson} \& {Green}(1998)}]{Peterson1998}
{Peterson} R.~C., {Green} E.~M., 1998, \apjl, 502, L39

\bibitem[{Piotto {et~al}\mbox{.}(2007)Piotto, Bedin, Anderson, King, Cassisi,
  Milone, Villanova, Pietrinferni, \& Renzini}]{Piotto2007}
Piotto G. {et~al.}, 2007, \apj, 661, L53

\bibitem[{Piotto {et~al}\mbox{.}(2005)Piotto, Villanova, Bedin, Gratton,
  Cassisi, Momany, Recio‐Blanco, Lucatello, Anderson, King, Pietrinferni, \&
  Carraro}]{Piotto2005}
---, 2005, \apj, 621, 777

\bibitem[{{Platais} {et~al}\mbox{.}(2011){Platais}, {Cudworth},
  {Kozhurina-Platais}, {McLaughlin}, {Meibom}, \& {Veillet}}]{Platais2011}
{Platais} I., {Cudworth} K.~M., {Kozhurina-Platais} V., {McLaughlin} D.~E.,
  {Meibom} S., {Veillet} C., 2011, \apjl, 733, L1

\bibitem[{{Ram{\'{\i}}rez} \& {Mel{\'e}ndez}(2005)}]{Ramirez2005}
{Ram{\'{\i}}rez} I., {Mel{\'e}ndez} J., 2005, \apj, 626, 446

\bibitem[{{Reimers}(1975{\natexlab{a}})}]{Reimers1975a}
{Reimers} D., 1975{\natexlab{a}}, M{\'e}moires of the Soci{\'e}t{\'e} Royale
  des Sciences de Li{\`e}ge, 8, 369

\bibitem[{{Reimers}(1975{\natexlab{b}})}]{Reimers1975b}
---, 1975{\natexlab{b}}, {Circumstellar envelopes and mass loss of red giant
  stars}, {Baschek, B., Kegel, W.~H., \& Traving, G.}, ed., p. 229

\bibitem[{{Renzini} \& {Fusi Pecci}(1988)}]{Renzini1988}
{Renzini} A., {Fusi Pecci} F., 1988, \araa, 26, 199

\bibitem[{{Schr{\"o}der} \& {Cuntz}(2005)}]{Schroder2005}
{Schr{\"o}der} K.-P., {Cuntz} M., 2005, \apjl, 630, L73

\bibitem[{{Schr{\"o}der} \& {Cuntz}(2007)}]{Schroder2007}
---, 2007, \aap, 465, 593

\bibitem[{{Scuflaire} {et~al}\mbox{.}(2008){Scuflaire}, {Th{\'e}ado},
  {Montalb{\'a}n}, {Miglio}, {Bourge}, {Godart}, {Thoul}, \&
  {Noels}}]{Scuflaire2008b}
{Scuflaire} R., {Th{\'e}ado} S., {Montalb{\'a}n} J., {Miglio} A., {Bourge}
  P.-O., {Godart} M., {Thoul} A., {Noels} A., 2008, \apss, 316, 83

\bibitem[{{Stello} {et~al}\mbox{.}(2010){Stello}, {Basu}, {Bruntt}, {Mosser},
  {Stevens}, {Brown}, {Christensen-Dalsgaard}, {Gilliland}, {Kjeldsen},
  {Arentoft}, {Ballot}, {Barban}, {Bedding}, {Chaplin}, {Elsworth},
  {Garc{\'{\i}}a}, {Goupil}, {Hekker}, {Huber}, {Mathur}, {Meibom},
  {Sangaralingam}, {Baldner}, {Belkacem}, {Biazzo}, {Brogaard}, {Su{\'a}rez},
  {D'Antona}, {Demarque}, {Esch}, {Gai}, {Grundahl}, {Lebreton}, {Jiang},
  {Jevtic}, {Karoff}, {Miglio}, {Molenda-{\.Z}akowicz}, {Montalb{\'a}n},
  {Noels}, {Roca Cort{\'e}s}, {Roxburgh}, {Serenelli}, {Silva Aguirre},
  {Sterken}, {Stine}, {Szab{\'o}}, {Weiss}, {Borucki}, {Koch}, \&
  {Jenkins}}]{Stello2010}
{Stello} D. {et~al.}, 2010, \apjl, 713, L182

\bibitem[{{Stello} {et~al}\mbox{.}(2008){Stello}, {Bruntt}, {Preston}, \&
  {Buzasi}}]{Stello2008}
{Stello} D., {Bruntt} H., {Preston} H., {Buzasi} D., 2008, \apjl, 674, L53

\bibitem[{{Stello} {et~al}\mbox{.}(2009){Stello}, {Chaplin}, {Basu},
  {Elsworth}, \& {Bedding}}]{Stello2009a}
{Stello} D., {Chaplin} W.~J., {Basu} S., {Elsworth} Y., {Bedding} T.~R., 2009,
  \mnras, 400, L80

\bibitem[{{Stello} {et~al}\mbox{.}(2011{\natexlab{a}}){Stello}, {Huber},
  {Kallinger}, {Basu}, {Mosser}, {Hekker}, {Mathur}, {Garc{\'{\i}}a},
  {Bedding}, {Kjeldsen}, {Gilliland}, {Verner}, {Chaplin}, {Benomar}, {Meibom},
  {Grundahl}, {Elsworth}, {Molenda-{\.Z}akowicz}, {Szab{\'o}},
  {Christensen-Dalsgaard}, {Tenenbaum}, {Twicken}, \& {Uddin}}]{Stello2011b}
{Stello} D. {et~al.}, 2011{\natexlab{a}}, \apjl, 737, L10

\bibitem[{{Stello} {et~al}\mbox{.}(2011{\natexlab{b}}){Stello}, {Meibom},
  {Gilliland}, {Grundahl}, {Hekker}, {Mosser}, {Kallinger}, {Mathur},
  {Garc{\'{\i}}a}, {Huber}, {Basu}, {Bedding}, {Brogaard}, {Chaplin},
  {Elsworth}, {Molenda-{\.Z}akowicz}, {Szab{\'o}}, {Still}, {Jenkins},
  {Christensen-Dalsgaard}, {Kjeldsen}, {Serenelli}, \& {Wohler}}]{Stello2011a}
---, 2011{\natexlab{b}}, \apj, 739, 13

\bibitem[{Stetson, Bruntt \& Grundahl(2003)Stetson, Bruntt, \&
  Grundahl}]{Stetson2003}
Stetson P.~B., Bruntt H., Grundahl F., 2003, Publications of the Astronomical
  Society of the Pacific, 115, 413

\bibitem[{Sweigart \& Gross(1976)}]{Sweigart1976}
Sweigart A.~V., Gross P.~G., 1976, The Astrophysical Journal Supplement Series,
  32, 367

\bibitem[{van Loon, Boyer \& McDonald(2008)van Loon, Boyer, \&
  McDonald}]{vanLoon2008}
van Loon J.~T., Boyer M.~L., McDonald I., 2008, \apj, 680, L49

\bibitem[{Ventura, D'Antona \& Mazzitelli(2008)Ventura, D'Antona, \&
  Mazzitelli}]{Ventura2008}
Ventura P., D'Antona F., Mazzitelli I., 2008, \apss, 316, 93

\bibitem[{von Hippel(2005)}]{vonhippel2005}
von Hippel T., 2005, \apj, 622, 565

\bibitem[{{White et al.}(2011)}]{White2011}
{White et al.}, 2011, \apj~submitted

\bibitem[{Wood(1979)}]{Wood1979}
Wood P.~R., 1979, \apj, 227, 220

\end{thebibliography}
\label{lastpage}
\end{document}